# The AI Wave and the Reinvention of Game Discovery: Oversupply, Structural Correction, and Agentic Player-Game Matching

**Brian Madanamootoo**
*University of Silicon Valley, California, USA*

## Abstract

AI-assisted production has sharply reduced the cost and team size required to ship a video game, producing a supply shock on open marketplaces. Recent estimates put Steam release volume at roughly sixty new titles per day, with median per-title revenue for a large share of releases falling below the platform's own submission fee [1]. This paper asks whether the resulting oversupply constitutes an emerging market crash or a structural correction, and what discovery infrastructure the market will require as a consequence.

We first quantify the 2010-2026 supply shock using a 93,073-title Steam metadata snapshot, a 200,000-interaction Steam user-behavior dataset, and itch.io catalog data, computing attention-concentration metrics directly (Gini coefficient of 0.96 over playtime, with the top 1 percent of titles absorbing 73.5 percent of total play hours), and we introduce generative asset-model release velocity on Hugging Face as a candidate leading indicator of production-cost decline. We then conduct a comparative-historical analysis against the 1983 North American video game crash, the closest documented case of supply-driven collapse in the medium's history, identifying which structural divergences (digital distribution, diversified incumbent revenue, and consolidation capital) redirect the present contraction toward concentration rather than collapse, drawing on incumbent evidence including Ubisoft's 2025-26 restructuring and its transfer of equity to Tencent-backed Vantage Studios. Third, we analyze Netflix Games, Xbox Game Pass, and the curated browser platform Poki as natural experiments in access-based distribution.

Building on the 1983 precedent, in which the market recovered through the insertion of a curation layer at the console gatekeeper level, we propose a scalable equivalent, an access model coupled with agentic player-persona matching, and report a computational cold-start pilot on real interaction data: a persona-proxy matcher achieves a 31.2 percent hit rate at rank 10 on titles held out entirely from training (95% bootstrap CI [28.6, 34.0]), a 2.7-fold lift over the random baseline, robust across 20 independent random splits and to a genre-tag ablation. The pilot tests a narrower claim than persona matching against all cold-start recommendation, specifically that persona-style profiles improve on genre-only content matching where pure interaction-only collaborative filtering cannot score unseen items at all. We close with a calibrated payout simulation, comparing four incentive structures on an identical revenue-sharing pool, showing that both payout design and matching quality independently move median per-title developer-side revenue, with simulated medians rising from roughly $250 under a status-quo-equivalent structure to $1,400-2,900 under redistributive structures at today's matching quality, and into four figures for all structures once matching quality improves; we define "sustain" against three explicit bars (fee recoup, improvement over baseline, and an illustrative viability anchor) rather than asserting it, and pre-register the corresponding human-subjects study.

### Research Questions

RQ1. To what extent has AI-assisted production shifted the volume and concentration of supply on open game marketplaces between 2019 and 2026?

RQ2. Do the structural conditions of the 2023-2026 contraction replicate the oversupply mechanism of the 1983 game market crash?

RQ3. Can agentic player-persona models combining psychological profiles with behavioral and social-distribution signals improve player-game matching over behavior-only collaborative filtering, particularly for cold-start titles?

RQ4. Under what payout and incentive structures does an access-based distribution model with agentic matching sustain median per-title developer-side revenue?

# 1. Introduction

Game development has been progressively democratized over the last decade by free and low-cost engines, cloud infrastructure, and, most recently, generative AI tools for art, dialogue, and code. Each wave lowered the barrier to shipping a game. The AI wave differs in degree rather than kind, but the degree matters: production tasks that once required a small team can now be executed by one person in a fraction of the time, at a fraction of the cost. The consequence is not a healthier market for small developers but a supply shock that has outpaced every existing discovery mechanism.

This paper treats that shock as a single causal chain with two divergent effects. At the low-effort end, falling production costs let marginal and AI-generated low-effort titles flood open storefronts. At the high-effort end, the same tools expand scope per development-dollar, raising the quality bar an ambitious title must clear to be noticed. Both effects increase the volume and variance of releases relative to the audience's finite attention, and neither effect is addressed by existing storefront algorithms, which reward pre-existing momentum (wishlists, reviews) rather than fit between a specific player and a specific game.

We ask whether this is best understood as an early-stage market crash, echoing the 1983 North American video game crash, or as a structural correction that concentrates revenue without collapsing it. We argue for the latter, but only after directly engaging the strongest published counter-position (Sections 2.2 and 4). We then argue that the historical resolution to 1983, a human curation layer inserted between an oversupplied catalog and the player, is the correct shape of solution but the wrong scale of mechanism for an era of tens of thousands of annual releases, and propose an agentic equivalent.

Our contribution is fourfold: (1) an empirical account of the supply shock computed from primary marketplace data, including the first attention-concentration metrics (Gini, Lorenz, top-share) reported for this question, extending a Steam-market literature that stops before the AI-era shock [35, 36, 37]; (2) a novel candidate leading-indicator metric, generative asset-model release velocity on Hugging Face [5]; (3) a directional inversion of the LLM persona-agent literature [28, 29, 30, 31, 32, 33], repurposed here as a matching mechanism and tested, on the narrow and precisely stated claim that persona-style profiles improve on genre-only content matching for titles pure interaction-only collaborative filtering cannot score at all, in a computational cold-start pilot with multi-seed and bootstrap robustness checks [4]; and (4) a calibrated payout simulation comparing four incentive structures on equal terms and connecting both structure and matching quality to median per-title developer-side revenue under access-based distribution, using Netflix Games, Xbox Game Pass, and Poki as the occupied, incomplete cells of the relevant design space [17, 18, 20].

# 2. Background and Related Work

## 2.1 Steam market studies and the saturation debate

Academic study of the Steam marketplace has tracked release volume, indie share, and developer counts through 2022, generally attributing discoverability declines to platform policy changes and the arrival of competing storefronts rather than to a production-cost shock [35, 37]. More recent topological work has modeled genre evolution as a structural phenomenon without addressing the AI-era supply increase [36]. A parallel industry literature disputes that Steam is meaningfully oversaturated at all, noting that the count of titles earning over ten thousand dollars in their first three months has risen year over year even as median outcomes remain poor, and attributing individual outcomes to quality, marketing, and execution rather than market-level saturation [38]. We engage this argument directly in Section 3.4: our position is that the market is not saturated in the sense of lacking room for good games, but is concentrating in the sense that an increasing share of attention and revenue accrues to a shrinking share of titles, which is a distinct and independently testable claim, and one we now test directly.

## 2.2 Historical crash and crisis literature

The games-history literature documents the 1983 crash in depth [7, 8, 9, 10]. The only academic treatment linking it to the present decade, published in ACM Games: Research and Practice in late 2025, applies a three-level macro, meso, micro framework to the 2023-2025 layoff wave and concludes that the industry's resilience, rooted in globalized production networks and diversified revenue streams, has thus far contained the crisis as a supply-side contraction rather than allowed it to become a systemic collapse [11]. We treat this as our direct predecessor and build on, rather than contradict, its finding, while extending its analysis with the AI-specific production-cost shock quantified in Study 1 and the 2025-2026 data window. Platform-studies literature on Steam's gamblified consumption model supplies the theoretical vocabulary for why platform operators are structurally disinclined to gate supply even when they can identify it [39].

### 2.3 Player modeling and game recommendation

The player-typology tradition, from Bartle's taxonomy [23] through Yee's motivational model [24] to Big Five personality studies in games, gives this paper its psychological grounding. On the recommender side, the closest prior work is Vu and Bezemer's method for improving indie discoverability by measuring similarity to already-successful titles [25], a momentum-reinforcing approach that, by construction, cannot help a title with no prior signal. A 2026 study clustering 771 players' attitudes toward game AI identified seven distinct profiles, offering a preliminary vocabulary for audience-sensitive design [26], and PsychoGAT demonstrates LLM interactive fiction as a psychological measurement instrument [27], a plausible substrate for persona elicitation without questionnaires.

### 2.4 LLM persona agents for recommender systems

Since 2024, LLM-based persona agents have become a fast-moving subfield of recommender-systems research: RecAgent [28] and Agent4Rec [29] established LLM-agent simulation of user-item interaction; PUB introduced a Big Five personality-driven simulator for recommender evaluation [30]; SimUSER [31], AlignUSER [32], and PersonaAct [33] have extended persona fidelity, world-model alignment, and multimodal persona synthesis, respectively; two surveys map the space [34]. Nearly all of this work uses persona agents to simulate users in order to evaluate a separate recommender system, in domains such as movies, news, e-commerce, and short-form video. To our knowledge, none applies persona agents as the matching mechanism itself, and none addresses games, a domain whose player-typology literature is unusually well developed relative to the domains this subfield has tested.

### 2.5 Access-model and bundling economics

Streaming-bundling economics from adjacent media establishes that access models redistribute rather than necessarily grow willingness to pay, and that subscription availability can depress full-price purchase intent elsewhere in a product's lifecycle [41]. We extend this literature into games via three live natural experiments (Section 5) and a calibrated payout simulation (Section 7).

## 3. Study 1: Quantifying the Supply Shock (RQ1)

### 3.1 Data and methodology

Study 1 uses four data sources, two of which we obtained and computed on directly. (a) A 93,073-title Steam metadata snapshot (appid, title, developers, genres, release date; coverage through October 2024), assembled by the open GameStatsHub pipeline from Steam's public endpoints [3], from which we compute annual release counts. (b) The Tamber Steam interaction dataset of 200,000 user-game events (12,393 users, 5,155 games, purchase and play events with hours) [4], from which we compute attention-concentration metrics. (c) The SteamDB release tracker and Indie Launch Lab annual reports [2], used to validate (a) and to extend the series to 2025. (d) The Hugging Face Hub API, sampled July 2026 across texture, 3D-asset, sprite, and related generation-model search terms, from which we compute asset-model release velocity [5]. Concentration is measured with the Gini coefficient over per-title totals, the associated Lorenz curves, and top-share statistics (share of total attention captured by the top 1 percent and top 10 percent of titles). All analysis code and intermediate tables are included in the reproducibility package (Appendix A).

### 3.2 Release volume: computed and validated

Figure 1 shows annual Steam releases computed directly from the metadata snapshot: 268 in 2010, 4,169 in 2016, 8,853 in 2020, and 16,939 in 2024. These computed values track the independent SteamDB series (Figure 2) within roughly 5 percent in overlapping years (for example, 7,486 versus 7,111 for 2018 and 16,939 versus 17,776 for 2024, the residual reflecting delisted titles absent from later snapshots), which validates the snapshot and lets us treat the SteamDB figure of 20,017 releases for 2025 [2] as the continuation of the same series. Release volume roughly doubled between 2020 and 2025, and industry reporting places early 2026 at approximately sixty releases per day, with close to half of releases receiving fewer than ten user reviews and a meaningful share failing to recoup the platform's $100 submission fee [1].

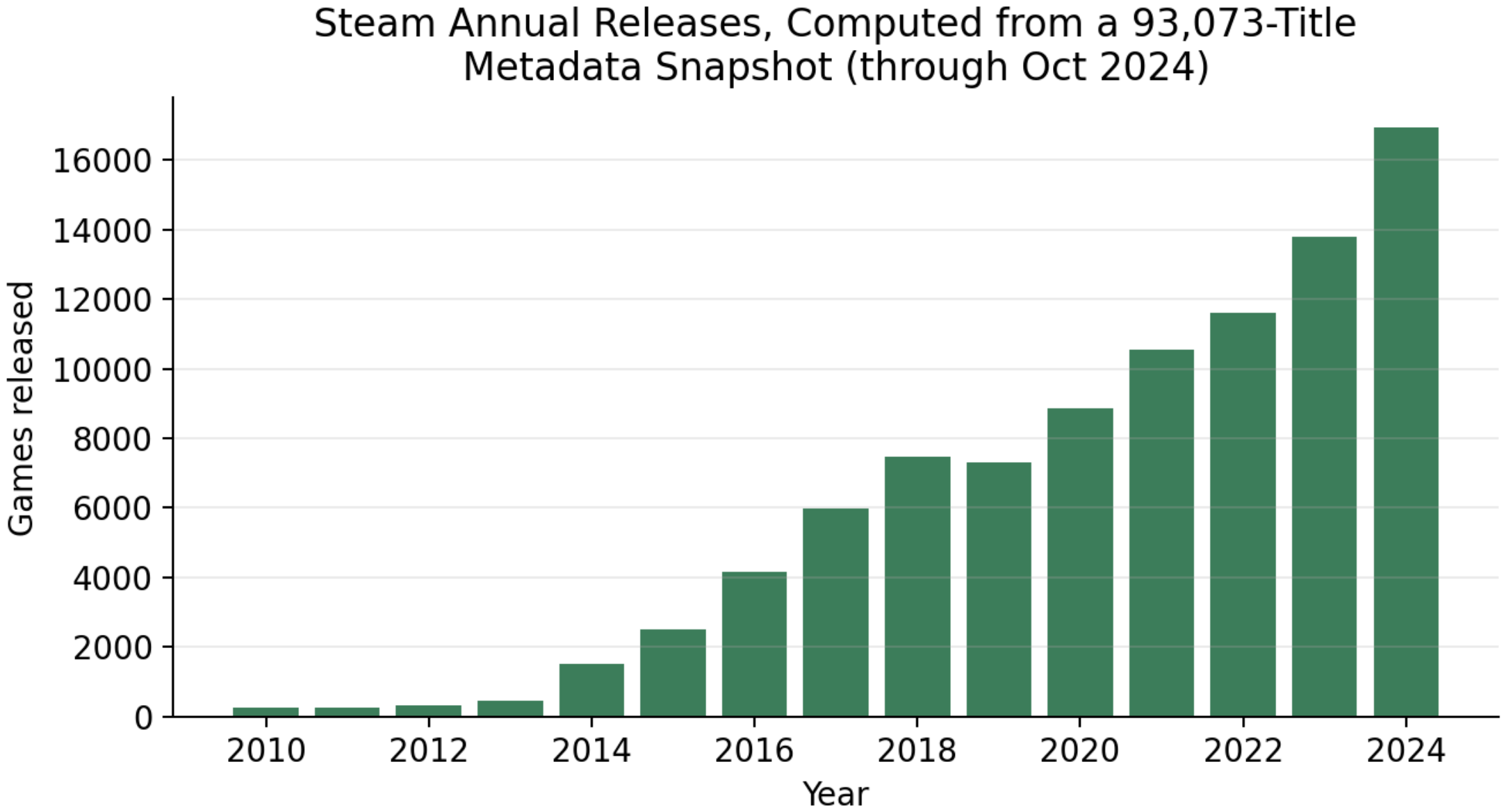

*Figure 1. Steam annual releases computed from the 93,073-title metadata snapshot (through October 2024), this paper's own computation [3].*

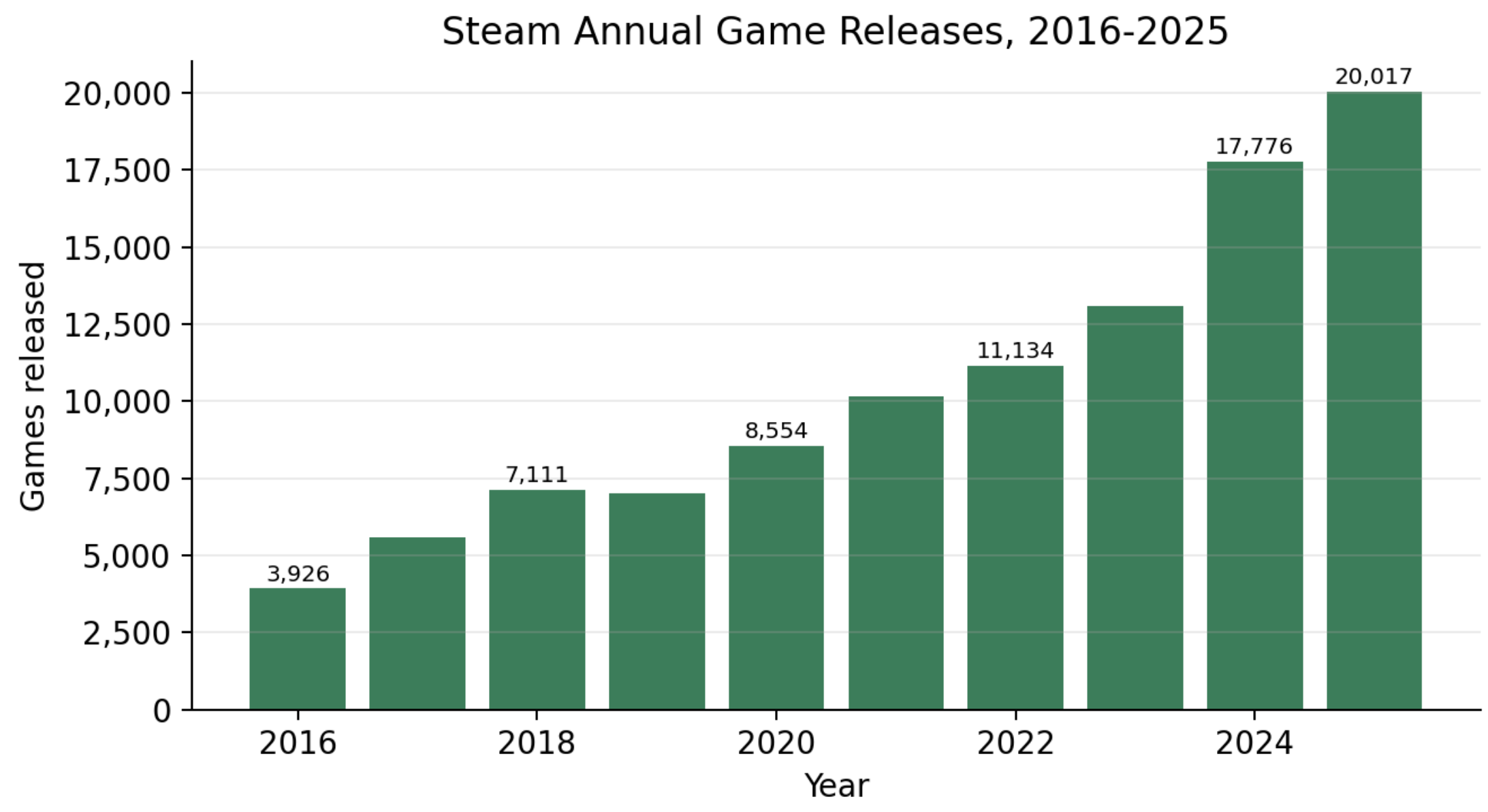

*Figure 2. Steam annual game releases, 2016-2025, from the SteamDB release tracker and Indie Launch Lab reports [2], shown for validation and 2025 extension of Figure 1.*

The concentration behind these totals is visible year over year in Steam's own review distribution: 49.8 percent of 2023 releases, 43.56 percent of 2024 releases, and 48.9 percent of 2025 releases received fewer than 10 user reviews, while only about 7.2 percent (2024) and 6.2 percent (2025) crossed 500 reviews, a threshold industry trackers use as a proxy for broad visibility (Figure 2b) [1, 43]. At the same time, platform-wide gross revenue kept growing across the same years, an estimated $15.0 billion in 2024 and $16.2 billion across the first eleven months of 2025 alone (Figure 2c) [1], the same divergence between rising aggregate revenue and a thinning per-title median that Section 3.4 formalizes as concentration.

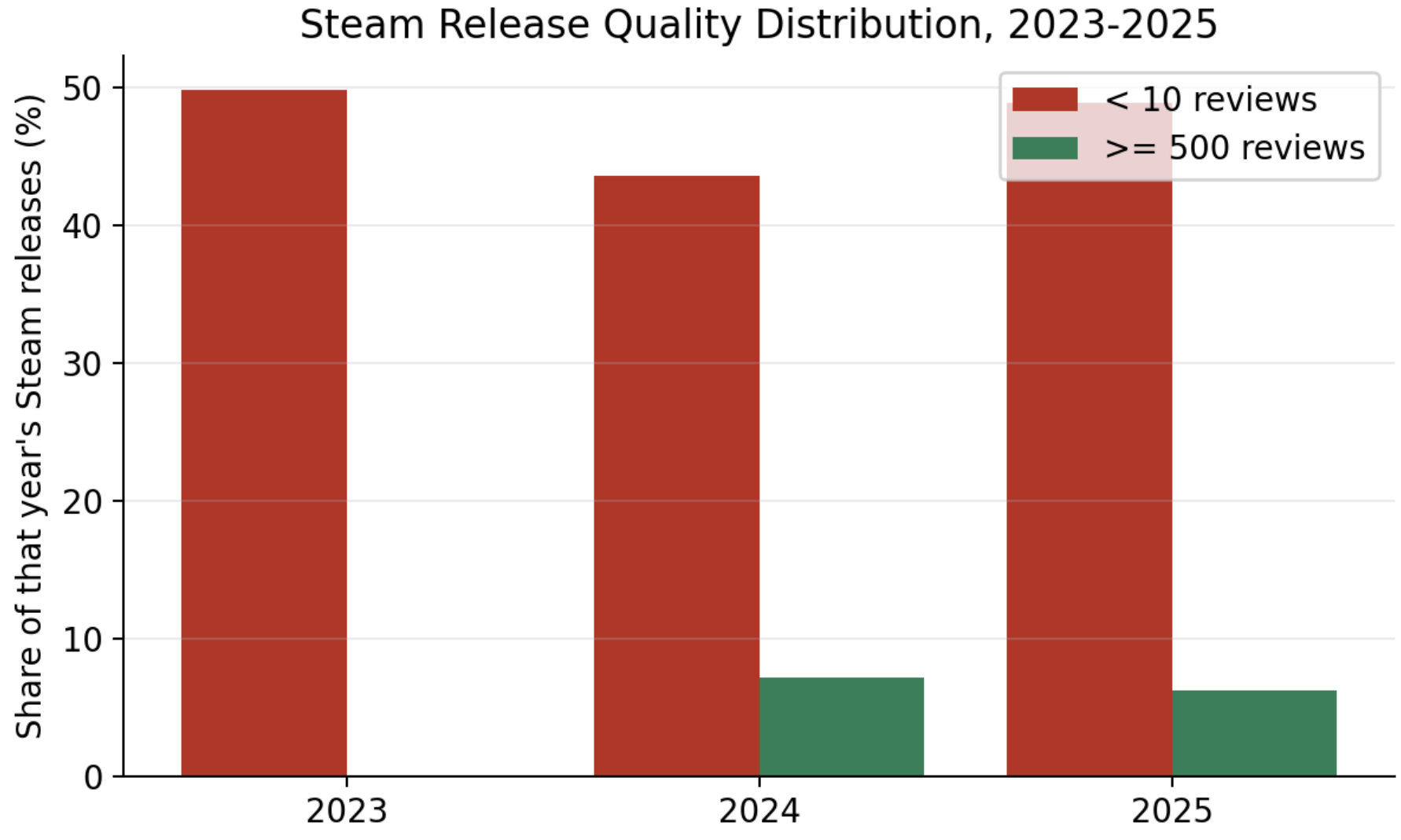

*Figure 2b. Share of Steam releases receiving fewer than 10 reviews versus 500 or more reviews, 2023-2025. Compiled from SteamDB-sourced industry analysis [1, 43].*

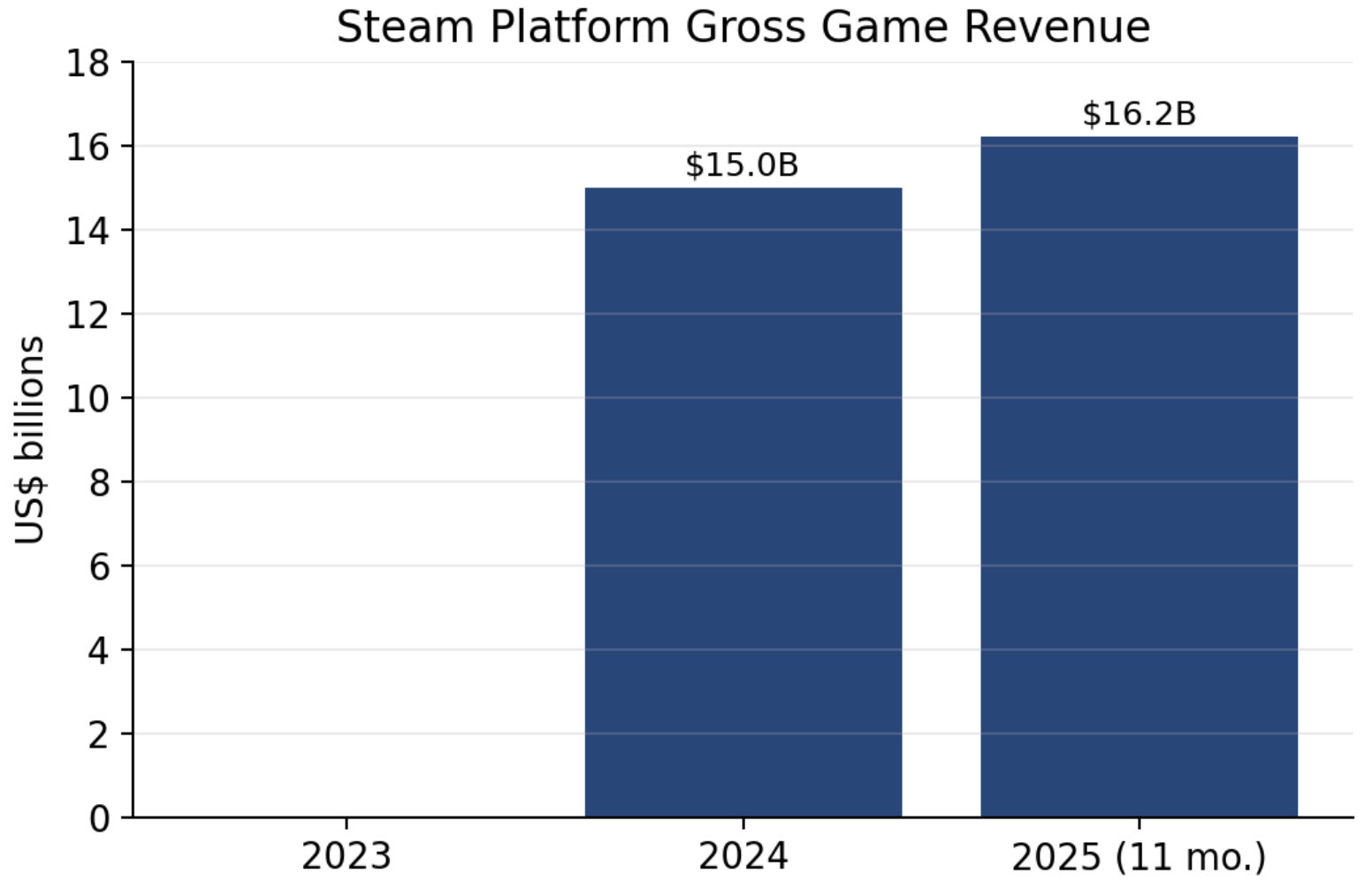

*Figure 2c. Steam platform gross game revenue, 2024 versus the first eleven months of 2025. Compiled from Alinea Analytics estimates as reported in industry press [1].*

itch.io provides the zero-friction limit case against which Steam's fee-gated model can be compared. As of mid-2026 the platform hosts over one million published products, including more than 557,000 entries originating from community game jams, all published with no upfront fee and no editorial review [42]. Release volume under near-zero friction is roughly an order of magnitude higher than Steam's, and per-product visibility is correspondingly lower, making itch.io the zero-friction end of the friction-to-discoverability curve used in Section 4.2.

## 3.3 Attention concentration: computed results

Using the interaction dataset [4], we compute the concentration of player attention across the 5,155 titles with recorded activity. The Gini coefficient over total playtime hours is 0.960; over purchase counts it is 0.782. The top 1 percent of titles absorb 73.5 percent of all recorded play hours, and the top 10 percent absorb 94.8 percent, while the median title accumulates roughly 15 total hours across the entire 12,393-user sample. Figure 3 shows the corresponding Lorenz curves. Two caveats govern interpretation. First, the interaction sample dates from the mid-2010s, before the AI-era supply shock, so these figures are best read as a lower bound on present concentration: attention was already near-maximally concentrated when annual supply was a quarter of today's, and every distributional statistic available for the later period (the sub-ten-review share, the median-revenue brackets [1]) points in the direction of further concentration, not less. Second, playtime concentration and revenue concentration are related but distinct; we treat playtime as the attention measure and rely on the secondary revenue statistics for the monetary claim, pending platform-scale revenue data.

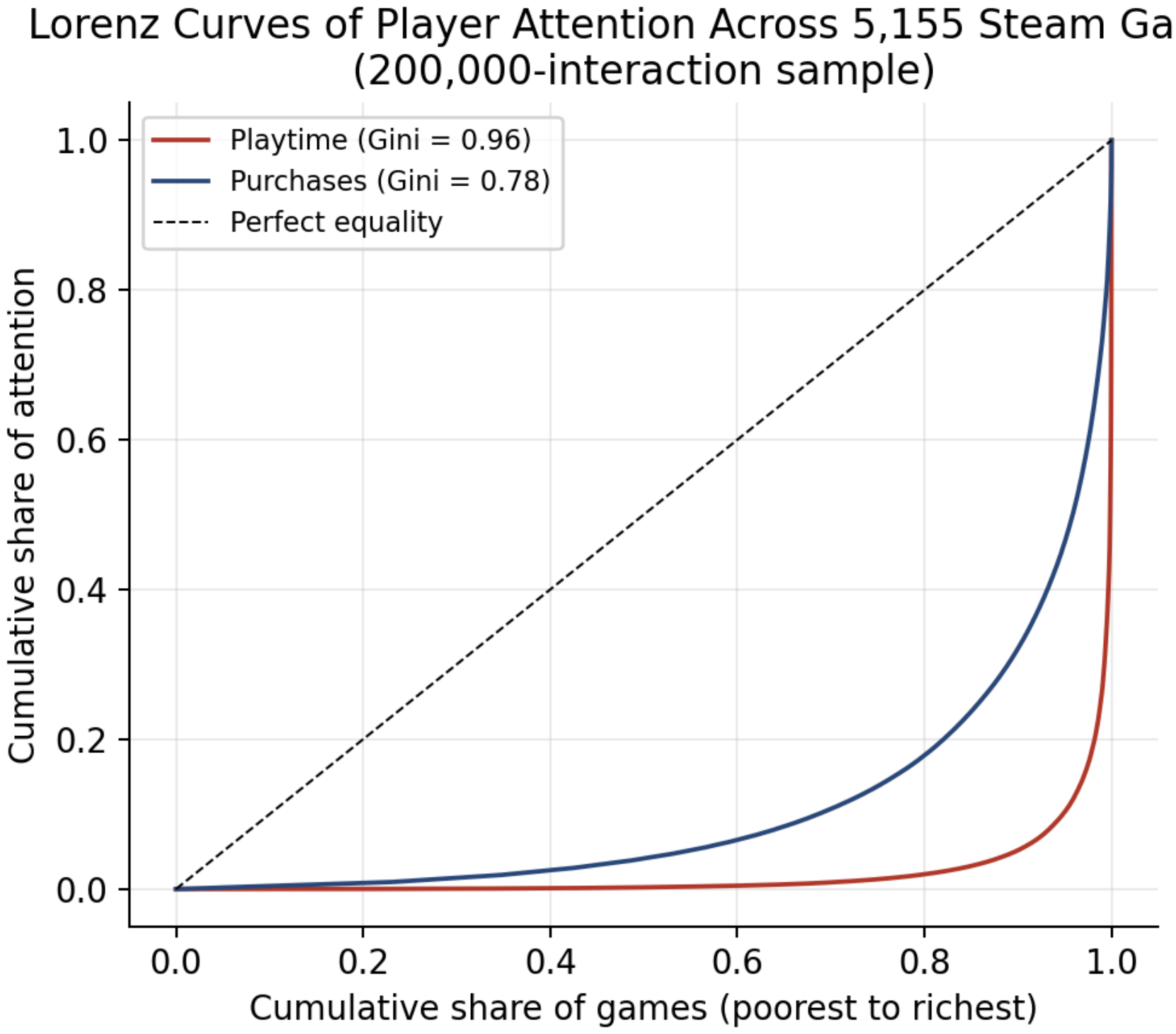

*Figure 3. Lorenz curves of player attention across 5,155 Steam titles in the 200,000-interaction sample [4], this paper's own computation. Gini = 0.96 (playtime), 0.78 (purchases).*

## 3.4 Concentration versus saturation

We distinguish two claims that are often conflated in industry commentary: that the market lacks room for new successful titles (saturation), and that an increasing share of attention and revenue accrues to a shrinking share of titles (concentration). Rising counts of individual breakout titles [38] are fully compatible with the concentration we measure, since both can be true simultaneously if the distribution's tail is thickening while its median is thinning. The computed Gini of 0.96 quantifies what the anti-saturation argument implicitly concedes: room at the top is not the same as viability at the median, and it is the median that the oversupply correction is damaging.

## 3.5 The Hugging Face leading-indicator candidate

Our sample of the Hugging Face Hub shows game-asset generation model releases growing from a single-digit count in 2023 to eighteen in 2024, fifty-six in 2025, and over one hundred thirty within the first seven months of 2026, with cumulative download counts rising roughly in step (Figure 4). This precedes and parallels the acceleration in marketplace release volume, which is consistent with a leading-indicator relationship, but we are explicit about what is and is not established: with four annual observations, no lag structure can be estimated, so we present this as a candidate indicator whose validation requires quarterly-resolution data on both series, specified as future work in Section 8.6. The sample is API-search-limited rather than a full Hub census, and a production-grade version of this metric should use the full Hub dataset with a fixed taxonomy of asset-generation tags.

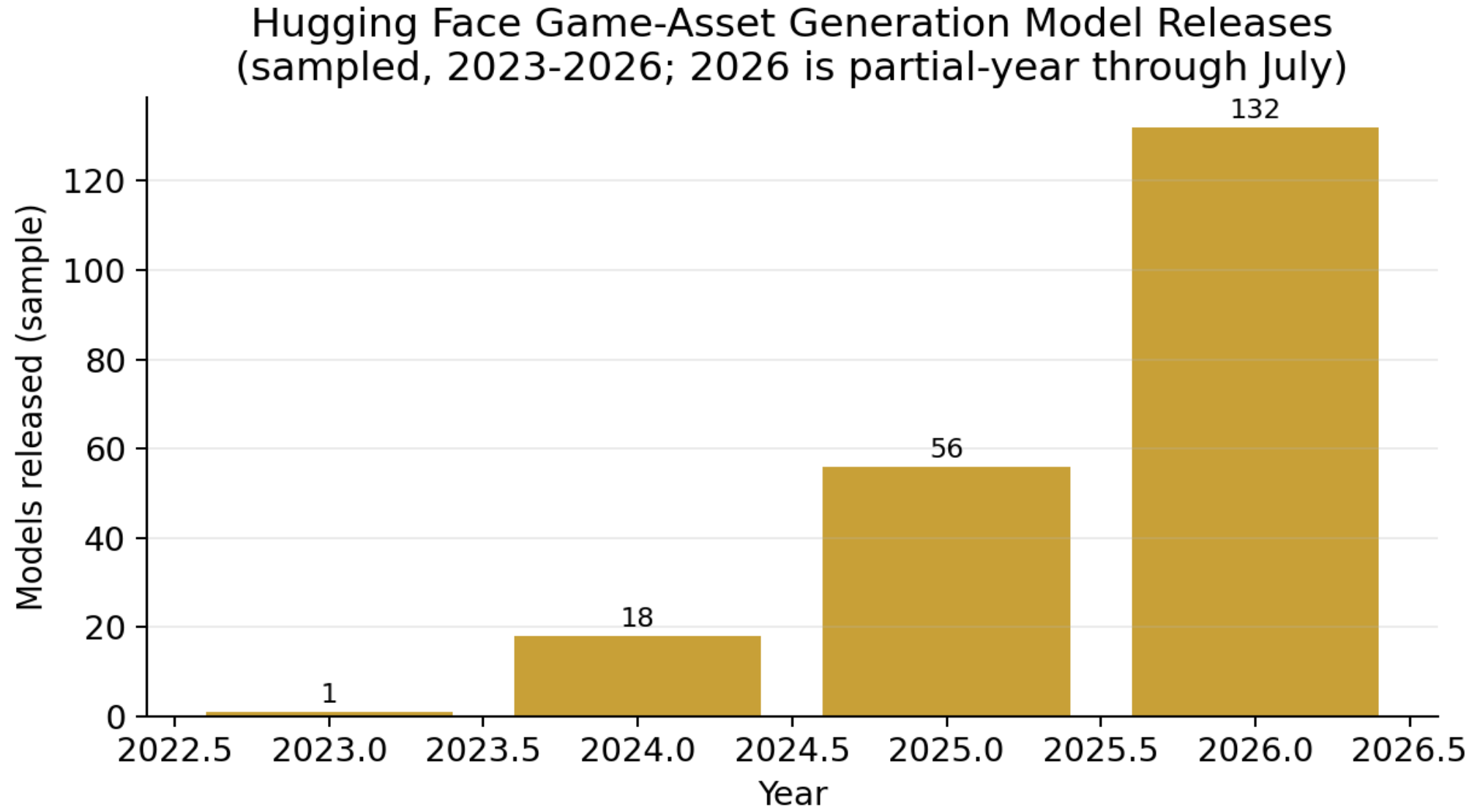

*Figure 4. Game-asset generation model releases sampled from the Hugging Face Hub API, by year (this paper's own data collection, July 2026) [5].*

## 3.6 Global overview: releases, attention, revenue, and review velocity

Figure 4b assembles the four quantities most often invoked separately in industry commentary, release volume, attention concentration, revenue, and the pace at which the market absorbs new titles, into a single reference figure. Panel A is the computed release series from Section 3.2. Panel B is the computed Lorenz curve from Section 3.3, restated here as the attention-concentration half of the picture. Panel C plots approximate global games-market revenue alongside Steam's own reported gross revenue where available, both essentially flat-to-growing across the contraction window, the single clearest visual rebuttal to a literal reading of oversupply as market collapse. Panel D reports a review-velocity proxy tracked by industry analysts, the number of copies sold per counted Steam review, which has fallen from roughly 93 in 2013 to roughly 32 by 2023 [43, 44]. We label this deliberately as a review-velocity proxy rather than a time-to-revenue measure: a falling ratio means reviews accumulate faster relative to units sold, which is consistent with either faster player engagement after purchase or changes in Valve's review-solicitation mechanics, and does not by itself establish that developers reach a revenue milestone sooner. No public time series tracks actual days-to-first-revenue-threshold across the full marketplace, and constructing one is listed as future work in Section 8.6; Panel D is included as the closest available longitudinal proxy rather than a substitute for that measurement.

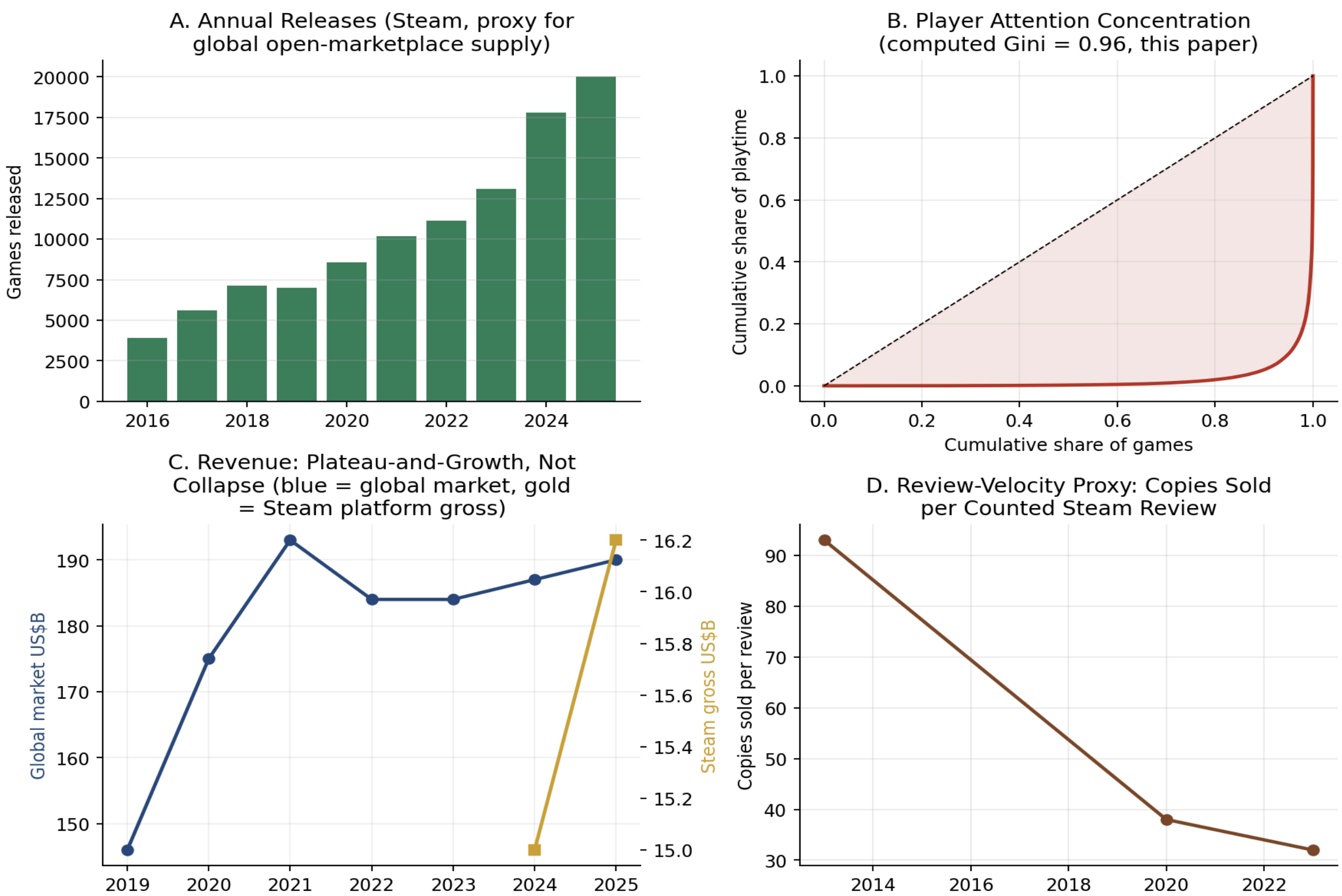

*Figure 4b. Global state of game releases, attention concentration, revenue, and review velocity. Panels A-B are this paper's own computations [3, 4]; Panel C compiles approximate global market revenue [40] and Steam's own reported gross revenue [1]; Panel D compiles the copies-sold-per-review benchmark from industry analytics [43, 44].*

# 4. Study 2: Correction or Crash, The 1983 Comparison (RQ2)

## 4.1 The 1983 mechanism

The 1983 North American crash is the medium's only fully documented case of a supply-driven collapse, and it remains the closest structural precedent for the present contraction, which is why it carries the comparative weight of this section rather than serving as a decorative historical aside. Between 1977 and 1982, home console sales grew from roughly 75 million dollars to over 3 billion dollars, and by 1982 an analyst at Goldman Sachs estimated that demand for video games had risen 100 percent year over year while manufacturing output had risen 175 percent, a supply/demand divergence created largely by unrestricted third-party publishing after Atari lost effective control over who could produce cartridges for its console [7]. The number of competing home console systems available in North America grew from about five in 1977 to more than fifteen by 1983 (Figure 5), fragmenting consumer attention and retail shelf space well before any single title's quality became the deciding factor [7, 9].

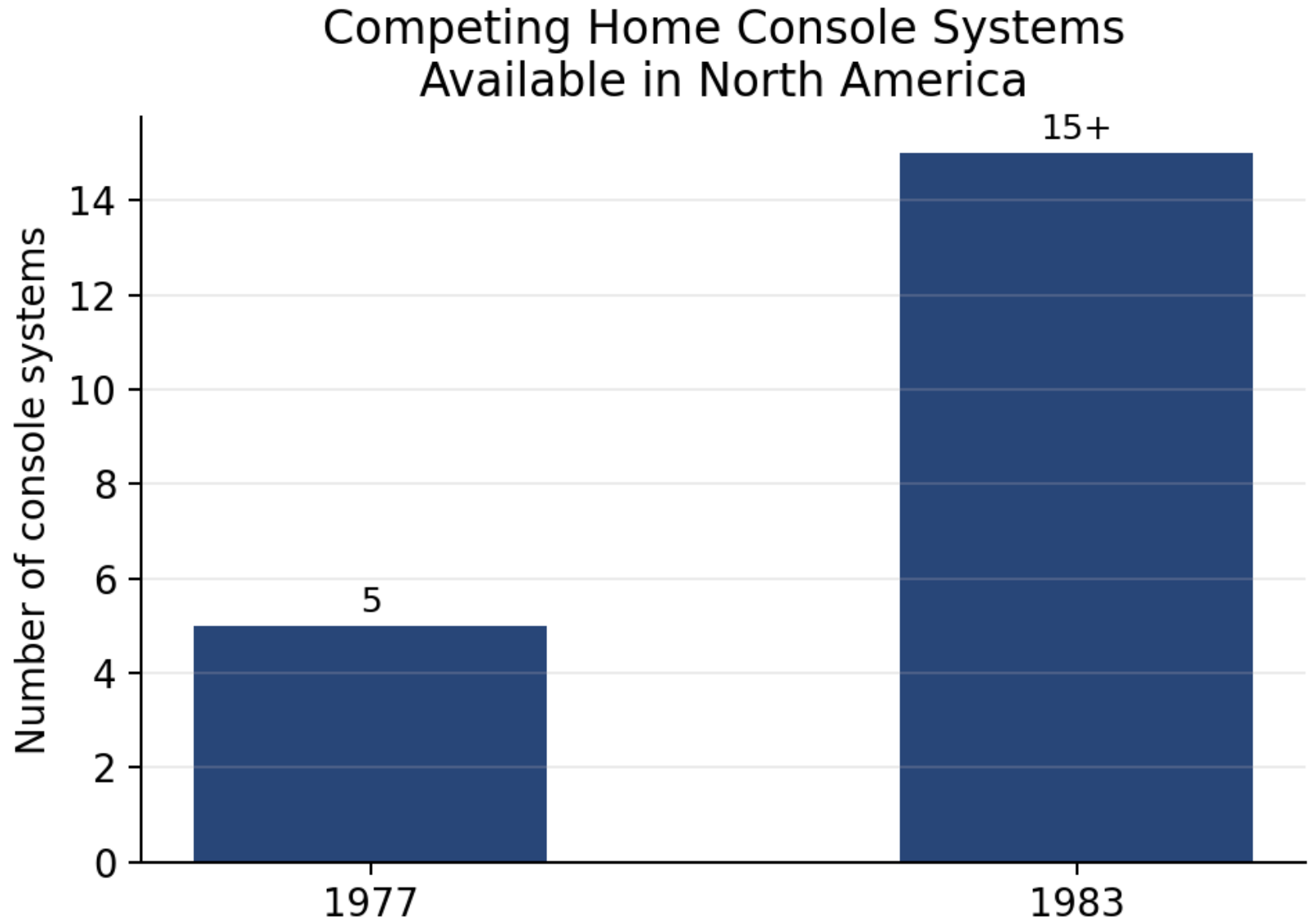

*Figure 5. Competing home console systems available in North America, 1977 versus 1983. Compiled from historical accounts of the pre-crash console market [7, 9].*

The quality collapse that followed is well documented at the level of individual products: Atari's 1982 conversion of Pac-Man sold only about 7 million of the 10 million cartridges manufactured despite a console install base of roughly 10 million, and the company's rushed Christmas 1982 release of E.T. the Extra-Terrestrial, produced in about five weeks, became a byword for the era's quality failures, with a large share of an estimated five million manufactured cartridges going unsold and, according to widely reported accounts, ultimately buried in a landfill [7, 9, 10]. Retailers responded rationally to a catalog they no longer trusted: unsold inventory was returned in bulk, shelf space for cartridges was cut, and toy retailers began treating console games as a passing fad rather than a durable category [7].

The financial and organizational damage was severe and concentrated. North American home video game revenue fell from a peak near 3.2 billion dollars in 1982 (equivalent to roughly 10.3 billion in 2025 dollars) to approximately 100 million dollars by 1985 (roughly 300 million in 2025 dollars), a decline of about 97 percent in nominal and real terms alike, before recovering past its prior peak by 1989 under Nintendo's post-crash market structure (Figure 6) [7, 8]. Atari alone reported a loss of roughly 356 million dollars and cut its workforce from a peak of about 10,000 employees in 1982 to roughly 3,000 by mid-1983 and to around 2,000 by 1984 (Figure 7), while Coleco, Imagic, Games by Apollo, US Games, and numerous smaller publishers exited the console business entirely or folded outright over the same window [10]. The market's recovery was led by Nintendo's insertion of a licensing and quality-control layer, the Seal of Quality program, between publishers and the retail shelf, restoring buyer trust in a catalog that had been discredited by the flood of low-quality titles described above [7, 9].

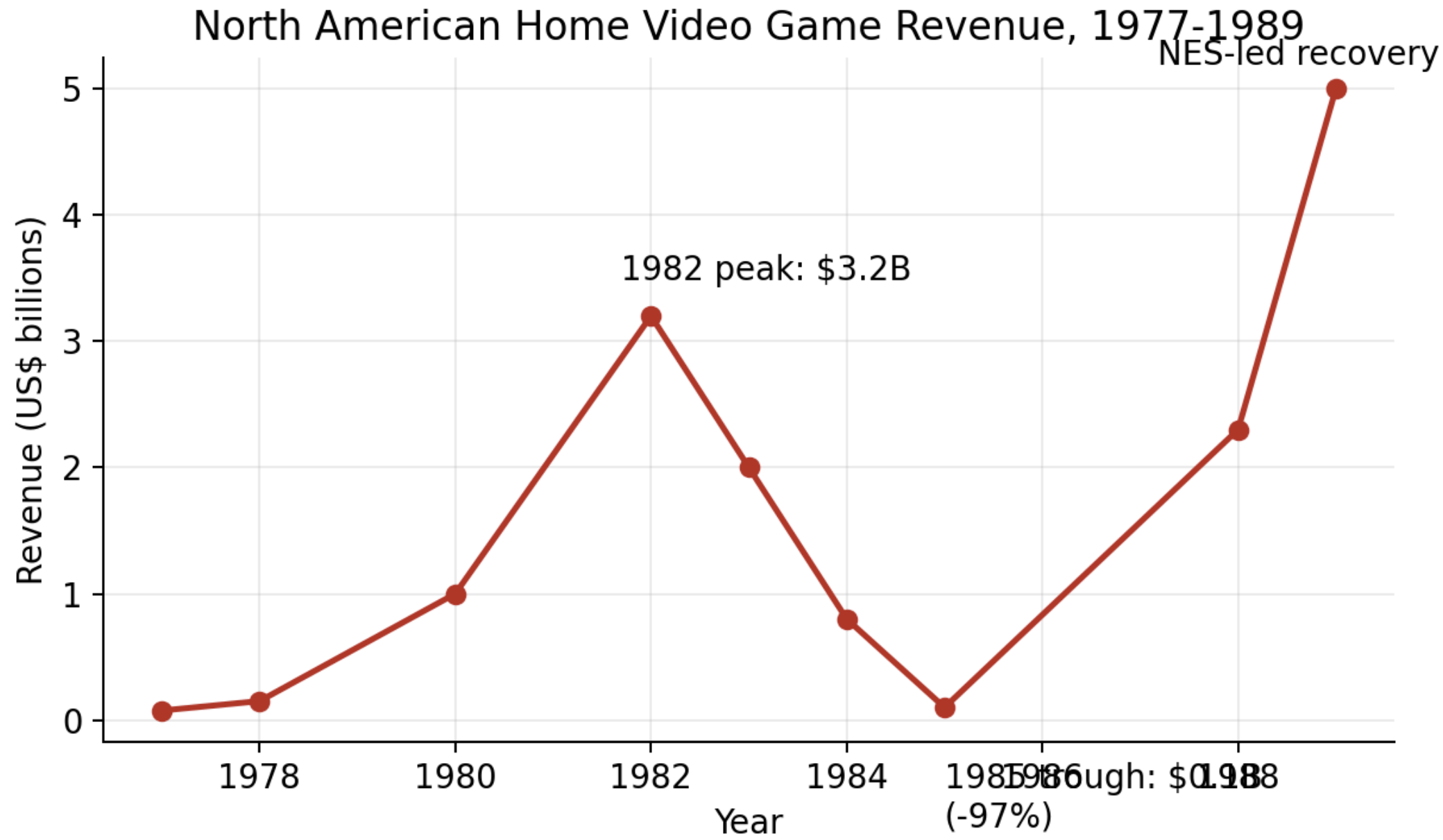

*Figure 6. North American home video game revenue, 1977-1989. Intermediate years between the documented 1982 peak and 1985 trough are illustrative interpolation, not independently sourced annual figures. Compiled from historical accounts [7, 8, 9].*

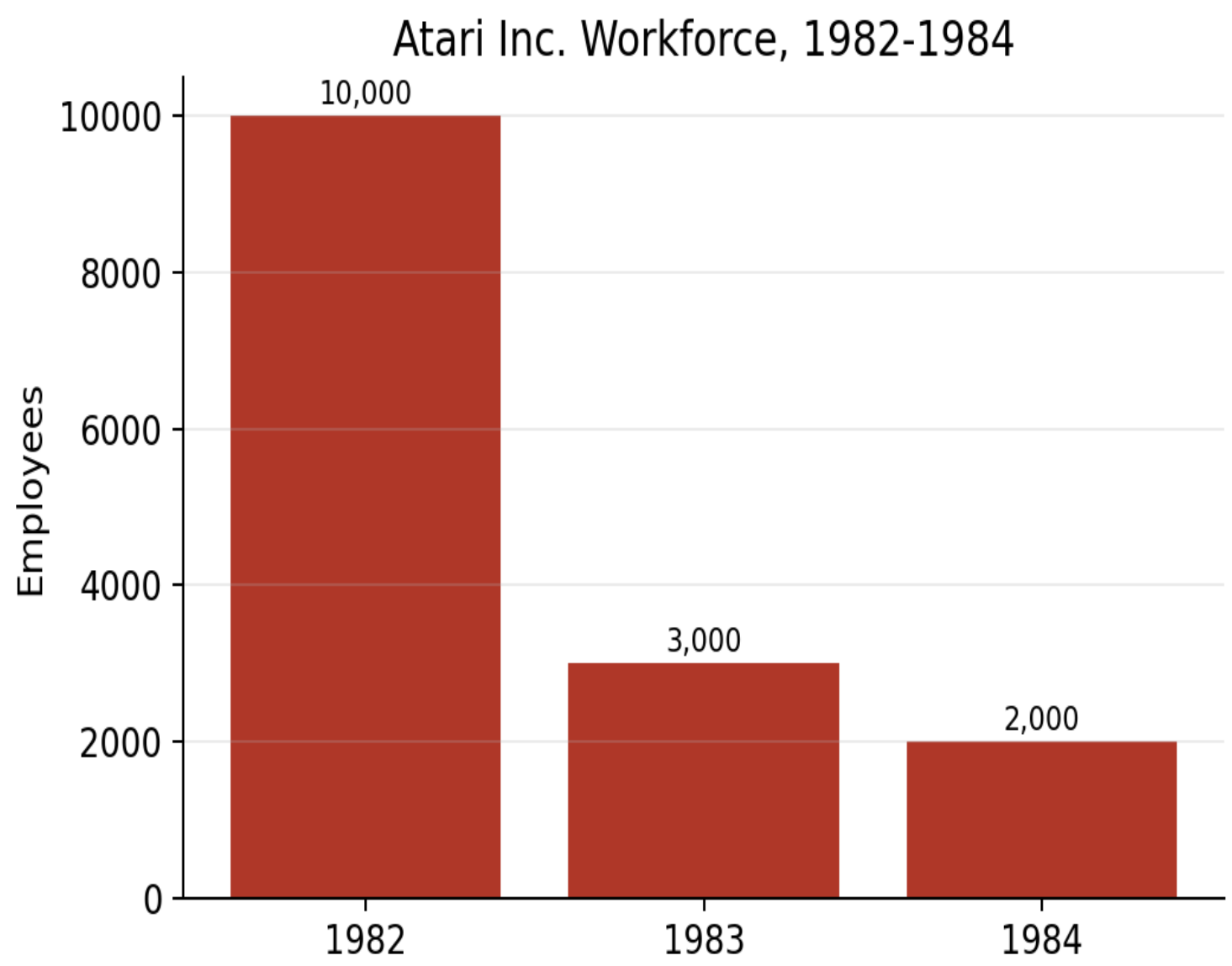

*Figure 7. Atari Inc. reported workforce, 1982-1984. Compiled from contemporaneous reporting [10].*

Academic treatment of 1983 beyond games-history specifically has generally used a "3Ds" framing (poor-quality software, market dilution, and platform proliferation, following the games-history literature summarized in Wolf's edited volume Before the Crash [9]) to explain the crash's proximate causes. The one existing academic work that connects 1983 directly to the present decade [11] extends that account with a three-level macro, meso, and micro framework applied to the 2023-2025 post-pandemic layoff wave, and concludes that the industry's resilience has thus far contained the present crisis as a supply-side contraction rather than allowed it to become a systemic collapse of 1983's scale. We treat this paper as our direct predecessor and build on its conclusion rather than contest it, while extending its analysis in two directions it does not cover: the AI-specific production-cost shock quantified in Study 1, and the 2025-2026 window, which includes Ubisoft's restructuring and is largely outside that paper's data collection period.

## 4.2 Structural mapping and divergences

Table 1 lays out the mapping explicitly, mechanism by mechanism, rather than leaving the 1983-to-present comparison at the level of broad analogy. Reading down the table, the first three rows are structural similarities, the same underlying mechanism operating in both periods, and the final four rows are the divergences that our analysis argues redirect the present contraction toward concentration rather than collapse.

| Dimension | 1983 | 2023-2026 |
|---|---|---|
| Core mechanism | Supply growth (175%) outpaces demand growth (100%) [7] | Production-cost collapse lets releases outpace attention (Study 1) |
| Quality-signal failure | No pre-release quality gate; retailers can't distinguish good cartridges from bad | No pre-release quality gate; storefronts can't distinguish good releases from AI-assisted low-effort ones |
| Platform response | None until Nintendo's post-crash entry; incumbents absorb the loss | Valve labels AI content (7,300+ titles by March 2026) rather than gating submissions [6] |
| Distribution channel | Physical retail shelf space, finite and channel-concentrated | Digital storefronts, effectively unlimited shelf space |
| Revenue diversification | Cartridge sales are the industry's primary and near-only revenue channel | Subscription, live-service, and advertising revenue diversify incumbent exposure |
| Failure resolution | Liquidation and exit (Coleco, Imagic, US Games, Games by Apollo) [10] | Recapitalization and restructuring (Ubisoft-Vantage [16], Sony/EA revenue-mix shifts [19]) |
| Curation resolution | Nintendo Seal of Quality: centralized licensing gate at console level [7, 9] | Not yet resolved; open storefronts have chosen labeling over gating (Section 4.1) |

*Table 1. Structural mapping between the 1983 crash and the 2023-2026 contraction, mechanism by mechanism.*

The single most consequential divergence is visible in aggregate revenue itself (Figure 8): 1982-1985 saw industry revenue collapse 97 percent, while 2019-2025 global games revenue has been flat to gently growing through the entire contraction, which is the clearest quantitative statement of the correction-not-crash thesis [40]. Reading the table's remaining divergence rows together: digital distribution removes the retail-shelf and physical-inventory failure points that amplified the original crash, since there is no finite shelf space to be wasted on unsold cartridges and no channel-wide retailer boycott of a discredited product category; diversified incumbent revenue means no single channel's collapse threatens total industry revenue as directly as cartridge sales did in 1983; and available consolidation capital allows distressed studios to be recapitalized rather than liquidated outright, a pattern documented concretely in the Ubisoft case below. The one row without a present-day resolution, curation, is precisely the gap this paper's proposed system (Section 6) is designed to fill.

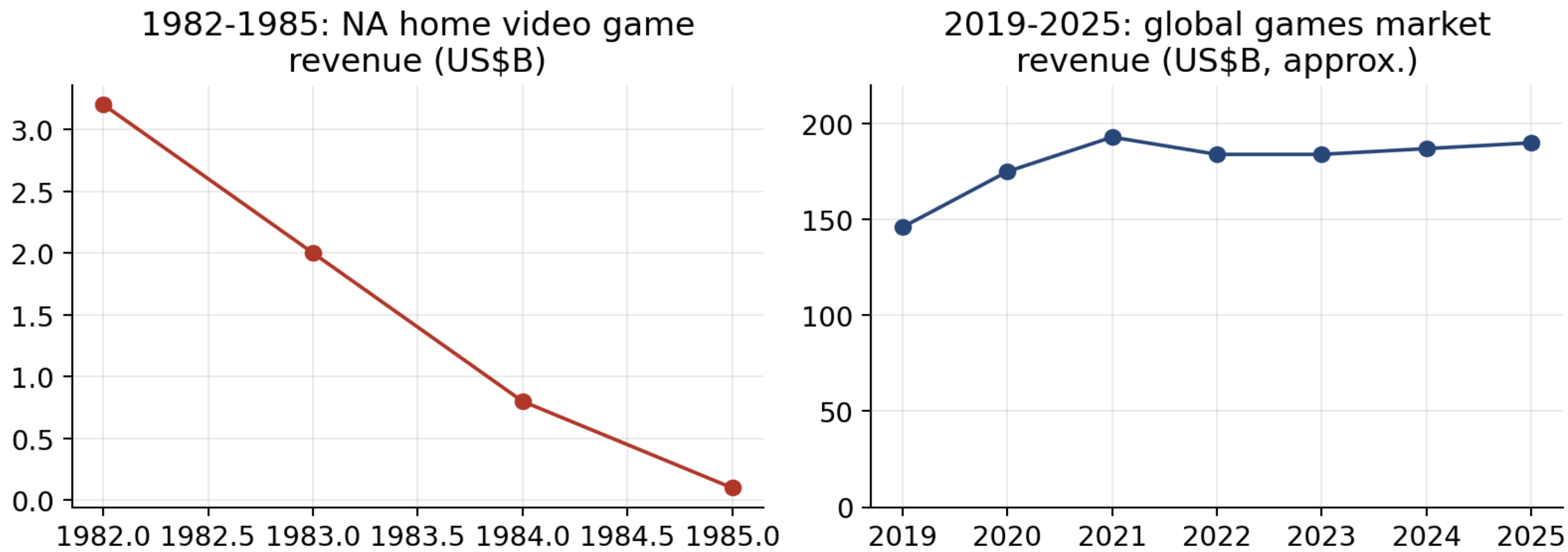

*Figure 8. Aggregate revenue trajectories: North American home video game revenue 1982-1985 (left) versus approximate global games market revenue 2019-2025 (right, compiled from Newzoo annual report figures [40]; approximate and shown for trajectory shape, not point precision).*

## 4.3 From industry-wide contraction to the Ubisoft case

Before turning to Ubisoft specifically, it is worth establishing that the contraction it exemplifies is industry-wide rather than company-specific, since a single distressed publisher would otherwise be weak evidence for a structural claim. Crowd-sourced and press-reported layoff tracking puts video game industry job losses at

approximately 8,500 in 2022, 10,500 in 2023, 14,600 in 2024, and a partial-year estimate in the 4,000 to 5,300 range for 2025 (Figure 9), for a cumulative total exceeding 40,000 positions since 2022 across companies including Microsoft Gaming, Sony Interactive Entertainment, Epic Games, Take-Two, Embracer Group, Unity, Riot Games, and Ubisoft [12]. A GDC-affiliated 2026 survey reported that roughly one third of US games-industry workers had been laid off at least once in the preceding two years [13]. This is the backdrop against which Ubisoft's specific results should be read, a severe but not idiosyncratic case within a sector-wide pattern, and the reason we treat it as an illustrative exhibit of the broader contraction rather than as an isolated corporate failure.

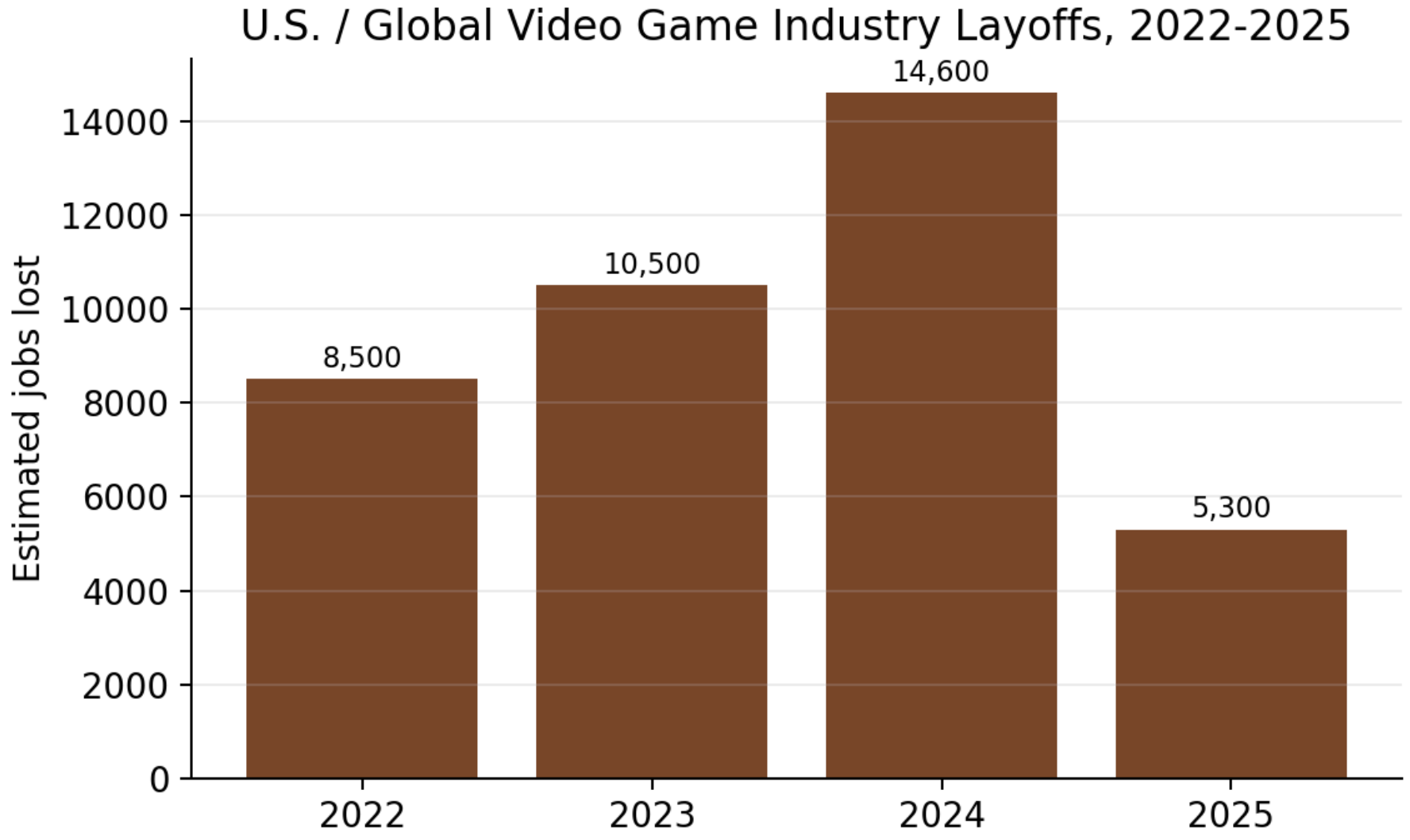

*Figure 9. Estimated video game industry layoffs, 2022-2025. Compiled from crowd-sourced industry layoff tracking and press aggregation [12].*

## 4.4 Incumbent evidence: the Ubisoft case

Ubisoft's fiscal year 2025-26 results, a record operating loss on declining net bookings and a stock price down roughly ninety percent from its 2018 peak, function as the paper's incumbent-scale exhibit within that wider contraction [15]. Annual closing share price fell from a high near $16 in 2018 to close to $2 by 2025 (Figure 10) [14], a decline that accelerated sharply around the January 2026 restructuring announcement and again after the FY2025-26 results were reported; on an intraday basis the stock has been reported as down as much as 93 to 96 percent from its mid-2018 peak [15]. Revenue and net income tell a complementary story (Figure 11): net income swung from modest positive figures in 2022 to a loss of roughly half a billion dollars in 2023, a partial recovery in 2024, and a return to loss in 2025, against a revenue base that itself contracted from roughly 2.3 billion dollars in FY2024 to an estimated 1.4 billion dollars on a trailing basis by FY2026 [14, 15].

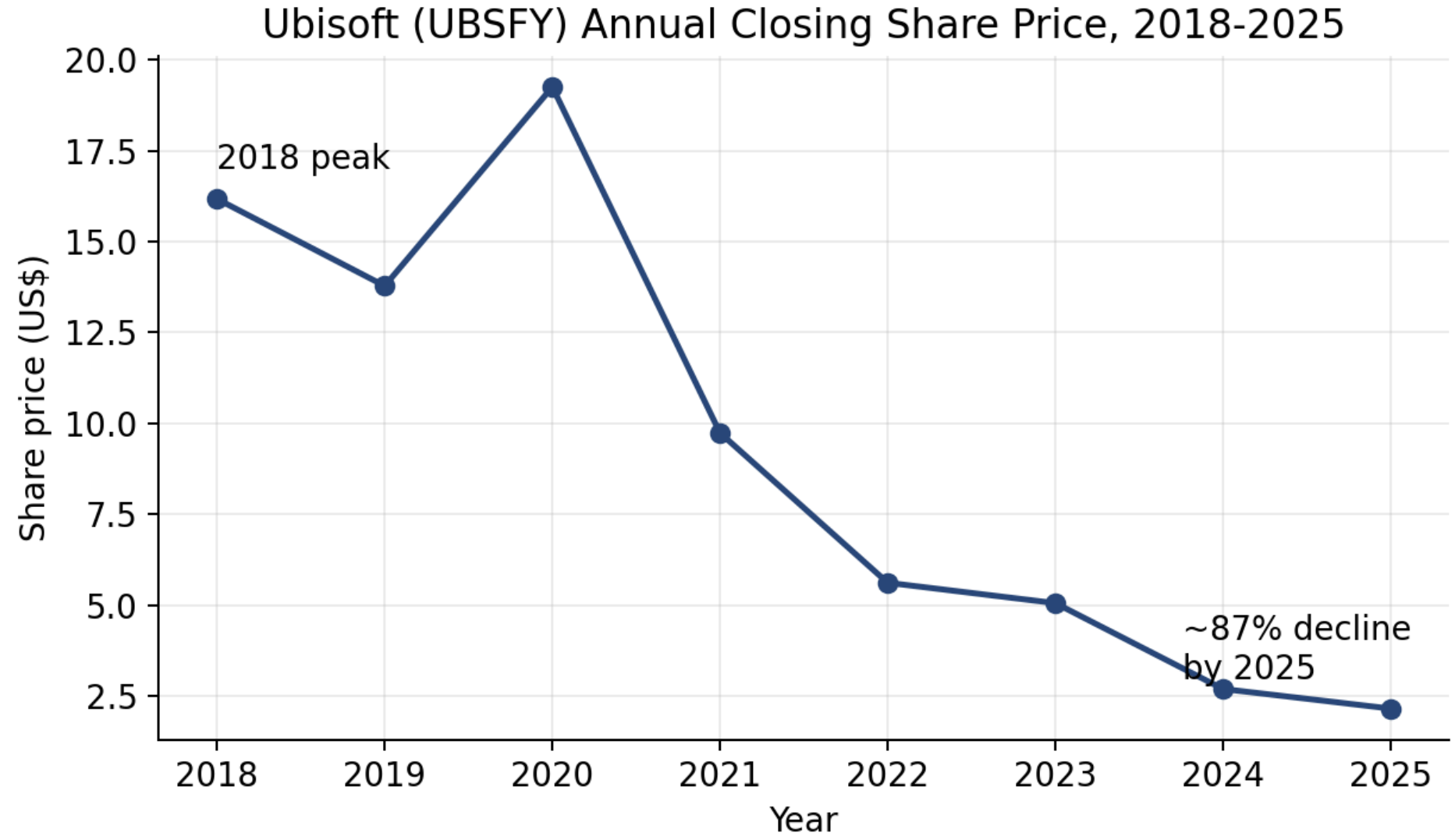

*Figure 10. Ubisoft (OTC: UBSFY) annual closing share price, 2018-2025, US dollars [14].*

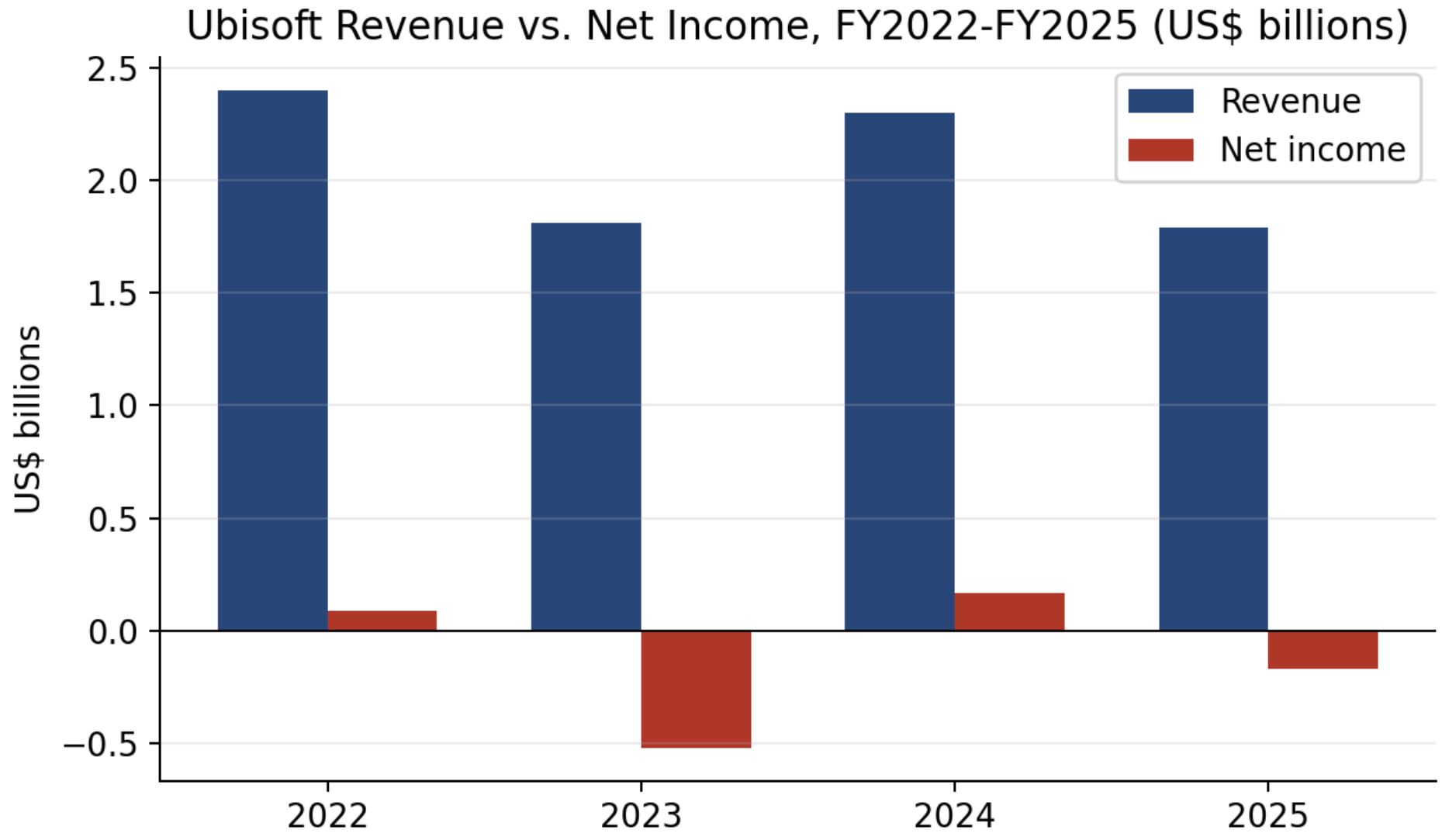

*Figure 11. Ubisoft annual revenue versus net income, FY2022-FY2025, US$ billions (approximate, compiled from company filings and financial-data aggregators [14, 15]). Bars below zero indicate a net loss year.*

The company's March 2025 transfer of its three largest franchises, Assassin's Creed, Far Cry, and Rainbow Six, into the Tencent-backed Vantage Studios, in exchange for approximately 1.16 billion euros for a 26.32 percent economic stake valuing the new entity above Ubisoft's own market capitalization at the time [16], is the clearest available 2025-26 instance of the consolidation-capital divergence identified in Table 1: distressed assets recapitalized and restructured rather than liquidated, a modern echo of Atari's 1984 sale to Jack Tramiel rather than of Intellivision's outright shutdown. Restructuring on the operational side has been comparably severe, roughly 1,200 positions cut across six waves with studio closures in Stockholm, Halifax, Winnipeg, and Belgrade, headcount down from approximately 20,729 in 2022 to roughly 16,600 by mid-2026, and a three-day strike by five French unions in February 2026 [15]. Sony's and EA's parallel shifts of revenue mix toward subscription and live-service models, discussed further in Section 5, supply corroborating, if less dramatic, evidence of the same industry-wide adaptation [19].

## 4.5 Verdict

The evidence assembled in this section, the historical mapping in Table 1, the aggregate-revenue divergence in Figure 8, the industry-wide layoff data in Figure 9, and the Ubisoft case in Figures 10 and 11, together support concentration and consolidation rather than 1983-scale collapse, consistent with and extending the conclusion of the existing ACM comparative analysis [11]. We treat the curation-layer resolution of 1983, the one row in Table 1 without a present-day equivalent, as the historically validated template for the present correction, while arguing in Sections 5 and 6 that human editorial curation at Nintendo's 1980s scale cannot reach the present release volume and that an algorithmic equivalent is required.

# 5. Natural Experiments in Access-Based Distribution

The proposed solution combines two components: subscription-style access, which removes per-title purchase risk, and agentic persona-based matching, which solves discovery at a scale human curation cannot reach. The live deployments reviewed here each test a subset of that combination, and none has tested both together, which frames the empty cell this paper proposes to fill.

## 5.1 Netflix Games: access without game-native matching

Despite being included at no marginal cost within a subscription exceeding 325 million paid memberships, Netflix Games reportedly reached less than one percent member engagement in 2023 [17]. The company subsequently abandoned planned AA and indie releases in favor of licensed IP tie-ins, and by its fourth-quarter 2025 earnings call had pivoted toward cloud-delivered, zero-literacy casual titles (Boggle, Pictionary, LEGO Party) explicitly framed as an engagement tool rather than a revenue source [17]. Catalog growth itself tracked this uncertainty: from 5 titles at the November 2021 launch to roughly 24 by the end of 2022 to 96 by mid-2024 (Figure 12b), followed by a wave of title removals beginning in December 2024 and continuing through 2026, including the closure of internal studio Boss Fight Entertainment [17]. We read this as evidence that bundled access alone, without a game-specific matching mechanism, does not convert a large subscriber base into meaningful engagement, and that the rational corporate response to that failure is retreat to pre-known IP, the least scalable available form of curation.

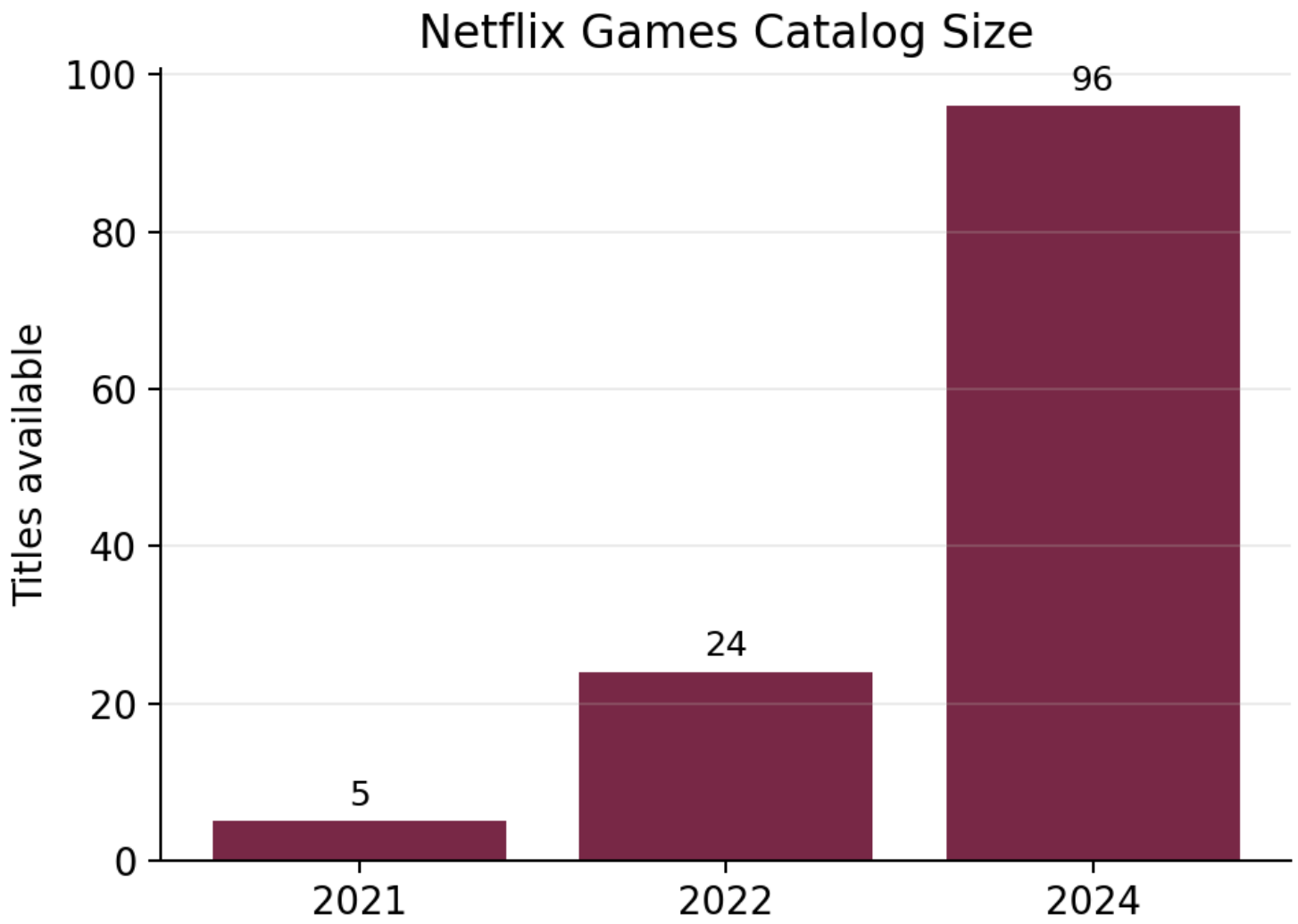

*Figure 12b. Netflix Games catalog size, 2021-2024, before the 2024-2026 title-removal wave. Compiled from press reporting [17].*

## 5.2 Xbox Game Pass: access with human curation

Game Pass reached a last officially confirmed thirty-four million subscribers in February 2024 against an internal projection, disclosed through litigation records in the Activision acquisition, of approximately seventy-seven million by 2026; subsequent reporting places the service around thirty million (Figure 12a) [18]. The full confirmed timeline, 10 million in April 2020, 15 million by September 2020, 18 million by January 2021, 25 million by January 2022, and 34 million by February 2024, shows steady growth slowing well before the price change discussed next [18]. An October 2025 price increase on the Ultimate tier from $19.99 to $29.99 reportedly drove meaningful subscriber loss, reversed by an April 2026 price reduction to $22.99 [18]. We treat this as a directly observable price-elasticity event, used to calibrate the economic model in Section 7. Some reporting attributes a substantial year-over-year increase in indie playtime to improved discovery features within the service, which, if independently verified against Microsoft's own disclosures rather than aggregator claims,

would constitute the closest existing evidence that discovery investment measurably improves indie engagement inside an access model; we flag this claim as requiring primary-source verification before use as evidence.

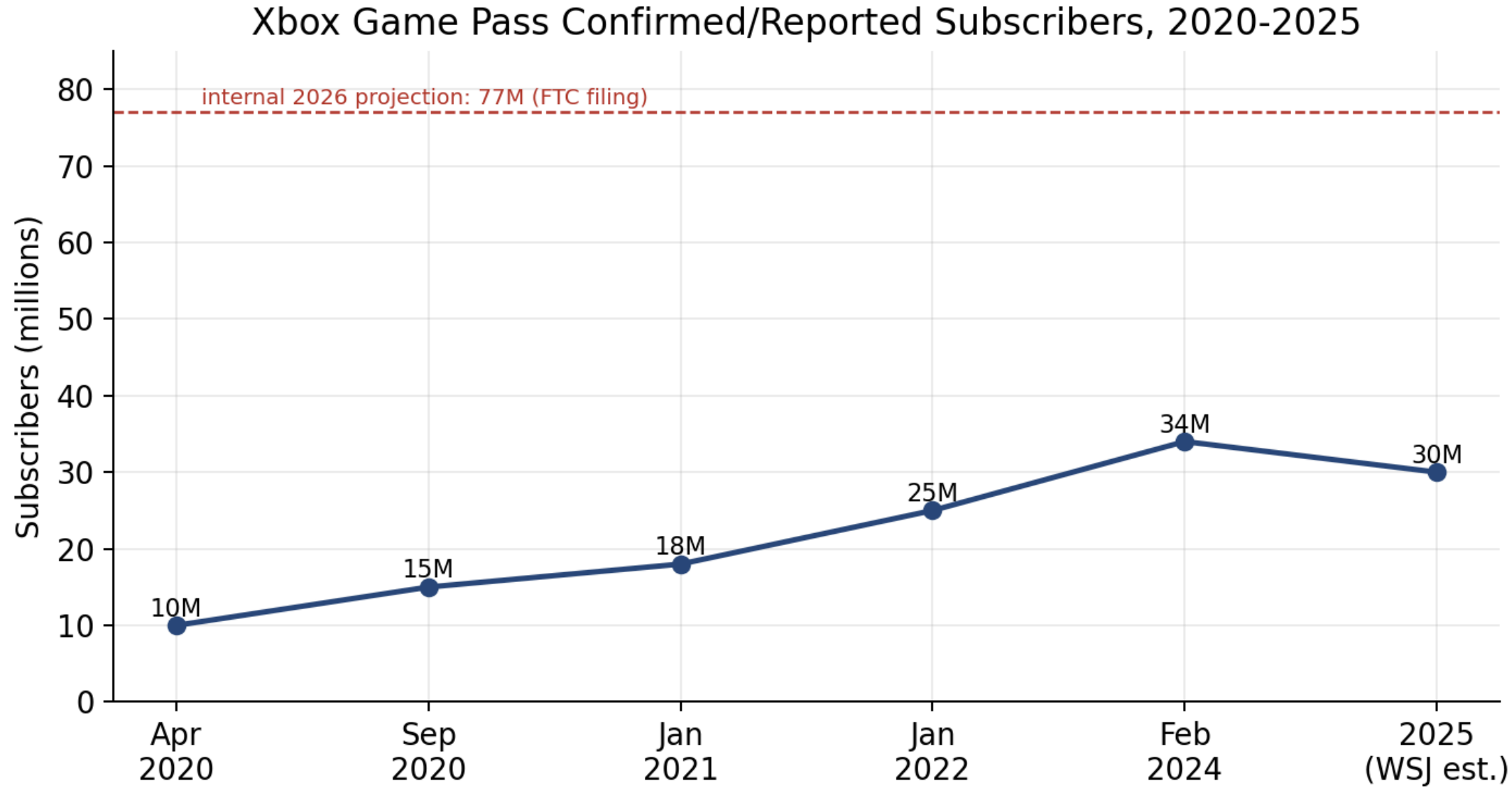

*Figure 12a. Xbox Game Pass confirmed and reported subscriber count, 2020-2025, against the internal 77-million 2026 projection disclosed in FTC v. Microsoft litigation records. Compiled from Microsoft disclosures and press reporting [18].*

## 5.3 Poki and browser-native curated distribution

A third, smaller-scale model is worth including precisely because it already combines two of the paper's proposed ingredients, human curation and algorithmic matching, at a scale between Netflix's mass bundling and Nintendo's 1980s licensing model. Poki, an Amsterdam-based browser-game platform founded in 2014, grew from tens of millions of monthly active players by 2016 to roughly 100 million monthly players and 1 billion monthly plays by mid-2025 (Figure 12), while curating a catalog of only around 1,000 to 1,500 titles from more than 600 developer studios, admitting new games through human editorial review rather than open submission, and reportedly releasing at a deliberately constrained rate of about one new title per day [20]. The company has never taken outside investment and operates with a team of roughly 65 people, which is itself informative: the platform's scale of player reach is now comparable to PlayStation Network's, achieved without the capital intensity of a console business [20].

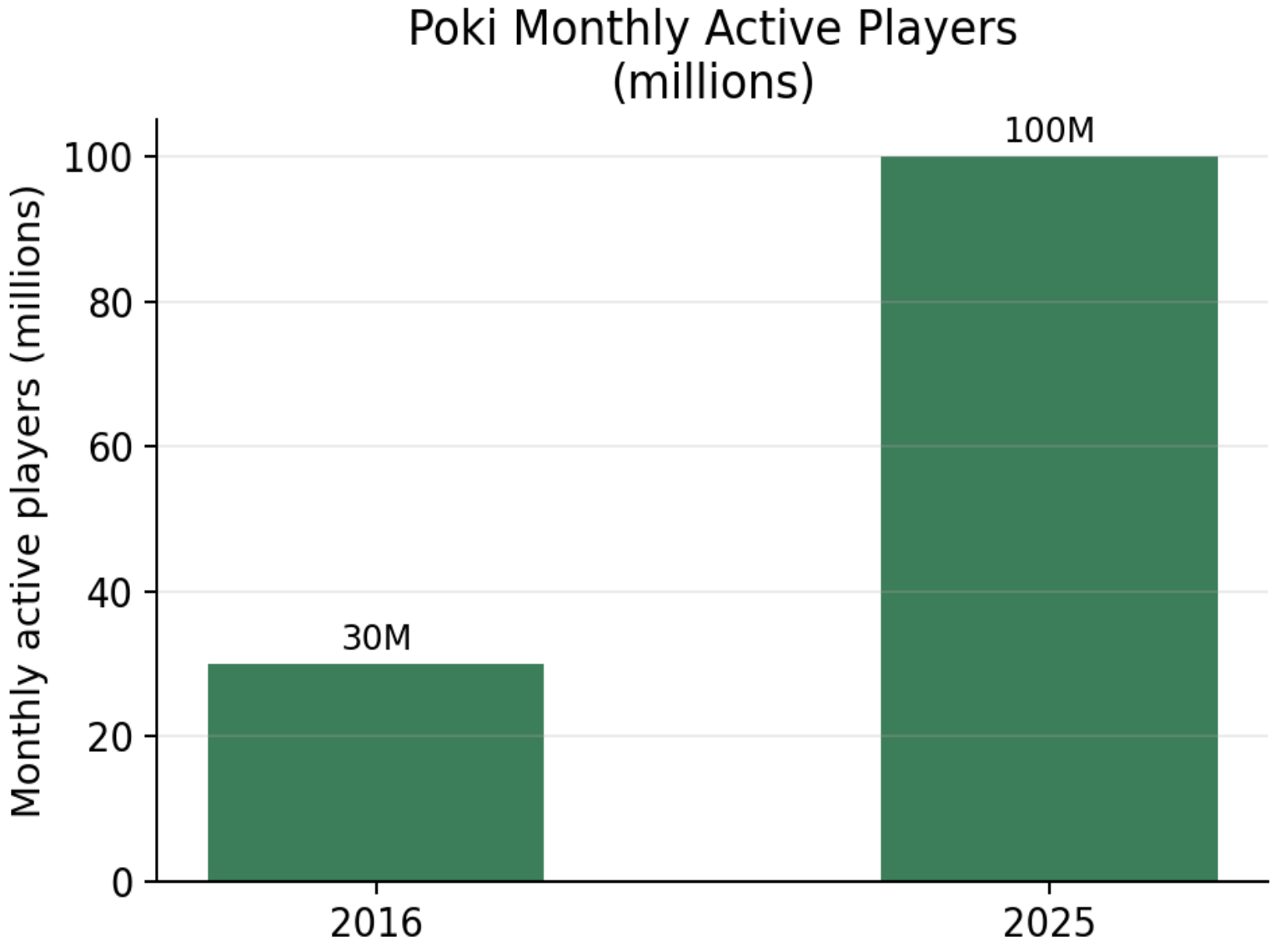

*Figure 12. Poki monthly active players, 2016 versus 2025 (millions). Compiled from company and press reporting [20].*

Distribution within the catalog is algorithmic, prioritizing average play time and conversion to play rather than acquisition spend, which the platform states is what lets a solo developer compete on comparable terms to a larger studio [20]. Monetization is entirely ad-based with a reported revenue share as high as fifty percent to developers, well above the roughly thirty percent norm on app-store style platforms, and top-performing studios are reported to have grown annual revenue from roughly fifty thousand dollars to as much as one million dollars over five years under this model (Figure 13) [20].

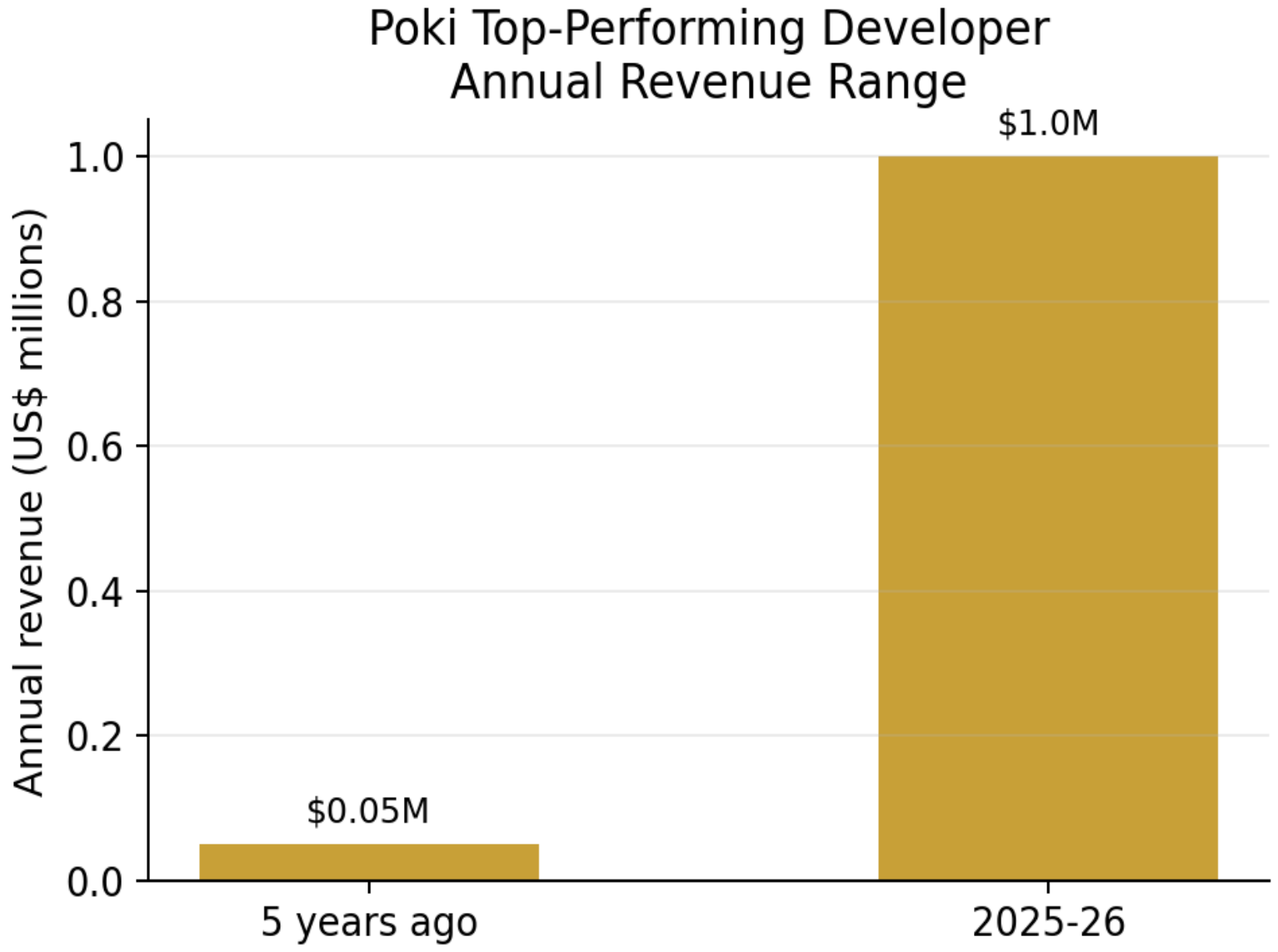

*Figure 13. Poki top-performing developer annual revenue range, five years ago versus 2025-26. Compiled from Dutch Games Association 2024 report figures as cited in press coverage [20].*

Poki is therefore a small, live instance of the historical resolution described in Section 4.1 and Table 1, human curation gating supply, paired with algorithmic within-catalog matching, at a scale (roughly a thousand titles) still small enough for human review to remain tractable. It is direct evidence that curation-plus-matching can work; the open question this paper addresses is whether an agentic, persona-based matching layer can preserve Poki's engagement-over-acquisition principle while scaling past the point where human editorial review of every submission remains feasible, which is the situation Steam and itch.io are already in.

## 5.4 Regional and platform-adjacent distribution: TapTap and Garena

Outside the Steam-centric Western PC market, two Asia-origin distribution models offer further comparative texture. TapTap, a China-origin mobile game discovery platform now used globally, hosts over 120,000 titles and more than 60,000 developer accounts, and organizes discovery primarily around a community review layer, more than four million user reviews read by upward of sixty million players, functioning as a crowd-curation mechanism rather than either open-storefront ranking or closed editorial review [21]. This is a third point on the curation-mechanism spectrum: neither Steam's momentum-weighted algorithm nor Poki's pre-publication human review, but large-scale post-publication crowd annotation.

Garena, the Singapore-based publisher and platform operator central to Southeast Asian mobile gaming, represents a different model again: rather than an open or semi-open catalog, Garena operates primarily as a first-party regional publisher and infrastructure provider (localized publishing, low-latency server architecture across ASEAN markets) for a comparatively small number of titles it chooses to bring to market, with its own Free Fire as the flagship example [22]. Garena's digital entertainment segment revenue peaked near $2.3 billion in 2021, fell to roughly $1.12 billion by 2023 as the pandemic-era player surge unwound, and recovered to approximately $1.9 billion in 2024 alongside a Free Fire resurgence and 661.8 million quarterly active users by Q1 2025 (Figure 12c) [22]. This is closer to traditional console-style publisher gatekeeping than to any of the marketplace models above, and it is included here as a reminder that not every regional market has followed the open-marketplace trajectory this paper otherwise focuses on; discovery is a non-issue for Garena's model because the catalog is small and centrally chosen, at the cost of the scale and developer diversity that Steam, itch.io, and Poki each provide in different degrees.

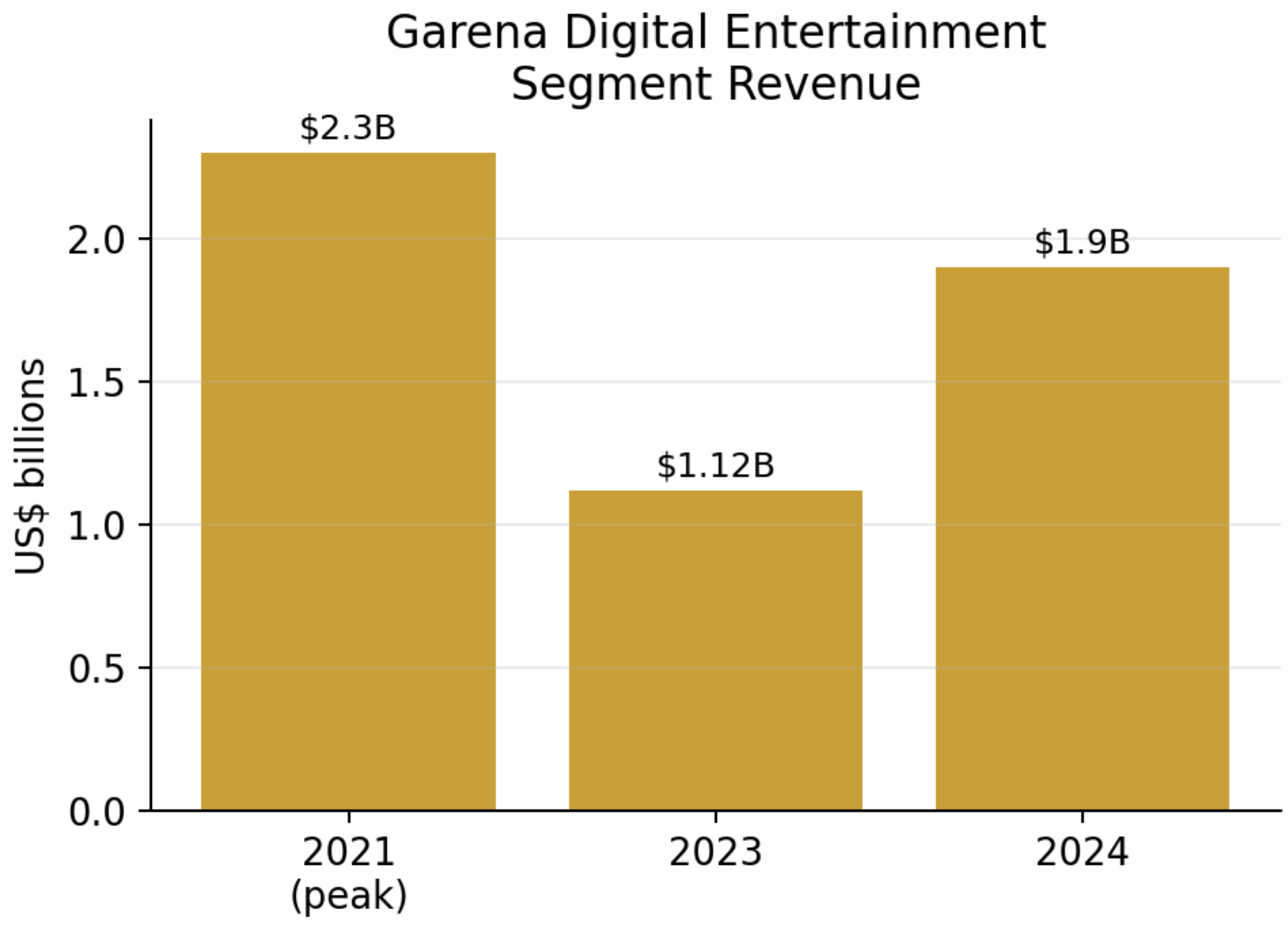

*Figure 12c. Garena digital entertainment segment revenue, 2021 peak versus 2023-2024. Compiled from Sea Limited earnings disclosures and financial press [22].*

### 5.5 The empty cell

None of the five deployments reviewed in this section tests access-based bundling combined with a psychologically grounded, agentic persona-matching layer, and none integrates distribution-native social signal (streaming clips, community reaction) as a matching input in the way Section 6 proposes. Poki's algorithmic layer comes closest, but operates on engagement and conversion signal within a pre-curated catalog small enough for human review, not on psychological persona modeling at open-marketplace scale. That combination, curation-equivalent matching that scales past human-review capacity, is the system proposed and piloted in Section 6.

## 6. Proposed System and Computational Pilot: Agentic Player-Game Matching (RQ3)

### 6.1 Architecture

The proposed system has three inputs feeding a single matching agent. (a) A psychological persona profile grounded in the Bartle, Yee, and Big Five traditions [23, 24], elicited either through a short instrument or through PsychoGAT-style interactive measurement [27], and maintained as an editable, player-visible object. (b) Behavioral signals: playtime, genre history, session patterns, abandonment points. (c) Social-distribution signals: clips, streams, and community reactions generated by distribution-native game design, which can exist before a title has any purchase or review history and therefore serve as the cold-start signal source. The matching agent is an LLM-based persona agent in the lineage of RecAgent and PUB [28, 30], but inverted in role: rather than simulating a user to evaluate a separate recommender, the persona agent scores candidate titles for fit against the maintained profile, producing a ranked slate with natural-language rationales that the player can inspect and correct, the correction itself becoming a profile update.

### 6.2 The directional inversion, precisely stated

Where the LLM persona-agent literature reviewed in Section 2.4 uses persona agents to simulate users for evaluating a separate recommender, we invert the direction: the persona agent is the matching mechanism. We state the cold-start claim precisely, since the strong version is not defensible and we do not make it. Modern recommender systems handle item cold start routinely through metadata embeddings, multimodal content models, two-tower architectures, graph features, franchise or creator side-information, contextual bandits, and hybrid collaborative-content approaches; persona-based matching is not the only, or even the first, cold-start technique in the field. The precise claim is narrower: pure interaction-only collaborative filtering, the specific incumbent mechanism examined in Section 3 and characteristic of momentum-based storefront ranking, cannot score a fully unseen item without side information, and the pilot below tests whether psychologically or persona-grounded profile information improves on the coarsest available side-information baseline, genre similarity, rather than testing persona matching against the full space of content-based cold-start methods. This is a repurposing of validated persona-agent machinery applied to a specific, narrow comparison, not a new technical claim about agent fidelity or a claim of exclusivity over cold-start recommendation generally.

## 6.3 Computational pilot: design

Before committing to a human-subjects study, we ran a computational pilot on real interaction data to answer the narrowest defensible version of RQ3: does a persona-style profile improve cold-start ranking over an unweighted popularity baseline and over chance, in a setting where the same profile has no purchase or interaction history to draw on for the held-out items? Data preparation addressed four integrity issues directly. First, the 200,000-event Steam interaction dataset [4] was filtered to remove 1,617 events (0.8 percent) tagged as downloadable content, soundtracks, art books, or cosmetic packs by name pattern, since these are not independently discoverable titles and would misrepresent the cold-start task; 4,982 base titles remain. Second, titles were joined to the genre metadata of the 93,073-title snapshot [3] by normalized title string; 65.4 percent (3,260 titles) matched. We checked this join for selection bias by comparing matched and unmatched titles on median playtime (14.3 hours matched versus 17.3 hours unmatched) and found no material skew, though mean playtime is somewhat higher among matched titles (998 versus 855 hours), consistent with modestly better metadata coverage for more-played titles. Third, 496 of 91,811 normalized title keys in the snapshot collide across more than one distinct game (common single-word titles such as "Alone" or "Echoes" recur across unrelated releases); collisions were resolved by taking the first-encountered genre assignment, a deterministic but non-adjudicated rule that we flag as a source of label noise in Section 6.5 rather than a resolved issue. Fourth, restricting to users with at least ten total played titles yields 1,144 eligible users; after also requiring at least five titles in the training partition and at least one title in the held-out partition (a small number of users are, by chance in a given random split, left with zero held-out titles), 1,116 users are evaluated under the main reported split, a loss of 28 users (2.4 percent) to this eligibility condition rather than to any modeling choice.

Protocol: titles are split 80/20 by uniform random sampling without replacement over the full title list, not chronologically and not stratified by genre or popularity; we report both a single main split (seed 42, used for the headline figures) and a 20-split robustness check below, since a single random split is not sufficient evidence of a stable effect size. The held-out 20 percent (499 titles under the main split) constitutes one fixed candidate set offered identically to every evaluated user, so no user receives a personalized candidate pool, only a personalized ranking over the same pool. Each user's persona proxy is computed only from their remaining (warm, non-held-out) titles as a log-playtime-weighted genre vector, a deliberately simple stand-in for the psychological profile of Section 6.1, and each method ranks the fixed 499 cold titles per user; a hit is recorded if any title the user actually played from that cold set appears in the method's top ten. All hit rates below are user-level means (macro-averages): each user contributes one binary hit/no-hit observation regardless of how many held-out titles they have, so heavy players are not over-weighted relative to light players. Comparators: (a) random ranking, the information floor; (b) the persona-proxy matcher (cosine similarity between the user's genre profile and each cold title's genre vector); (c) a popularity oracle ranking cold titles by their true held-out popularity, which no deployed momentum-based system could observe for a genuinely unseen title and which therefore serves as an unobtainable ceiling rather than a baseline a system could target.

## 6.4 Computational pilot: results

Evaluated over 1,116 users under the main split, the persona-proxy matcher achieves a hit rate at rank 10 of 31.2 percent (95 percent bootstrap confidence interval [28.6, 34.0], 3,000 resamples), against 11.4 percent for random ranking, a 2.7-fold lift, with the popularity oracle at 86.7 percent marking the ceiling that momentum signal would provide if it were observable at cold start, which it is not (Figure 14). The number of relevant (held-out, actually-played) titles per evaluated user has mean 6.7 and median 4 (interquartile range 3 to 8, range 1 to 82), so the typical user's hit/no-hit outcome reflects a small handful of relevant items against a 499-item candidate pool, not a dense signal. This matters for interpreting the random baseline: given the observed per-user relevant-item counts, the analytically expected chance hit rate at rank 10 against a 499-item pool is 12.0 percent, closely matching the 11.4 to 11.7 percent observed across the main split and the robustness runs below, confirming the random baseline behaves as combinatorics predicts rather than reflecting an implementation artifact.

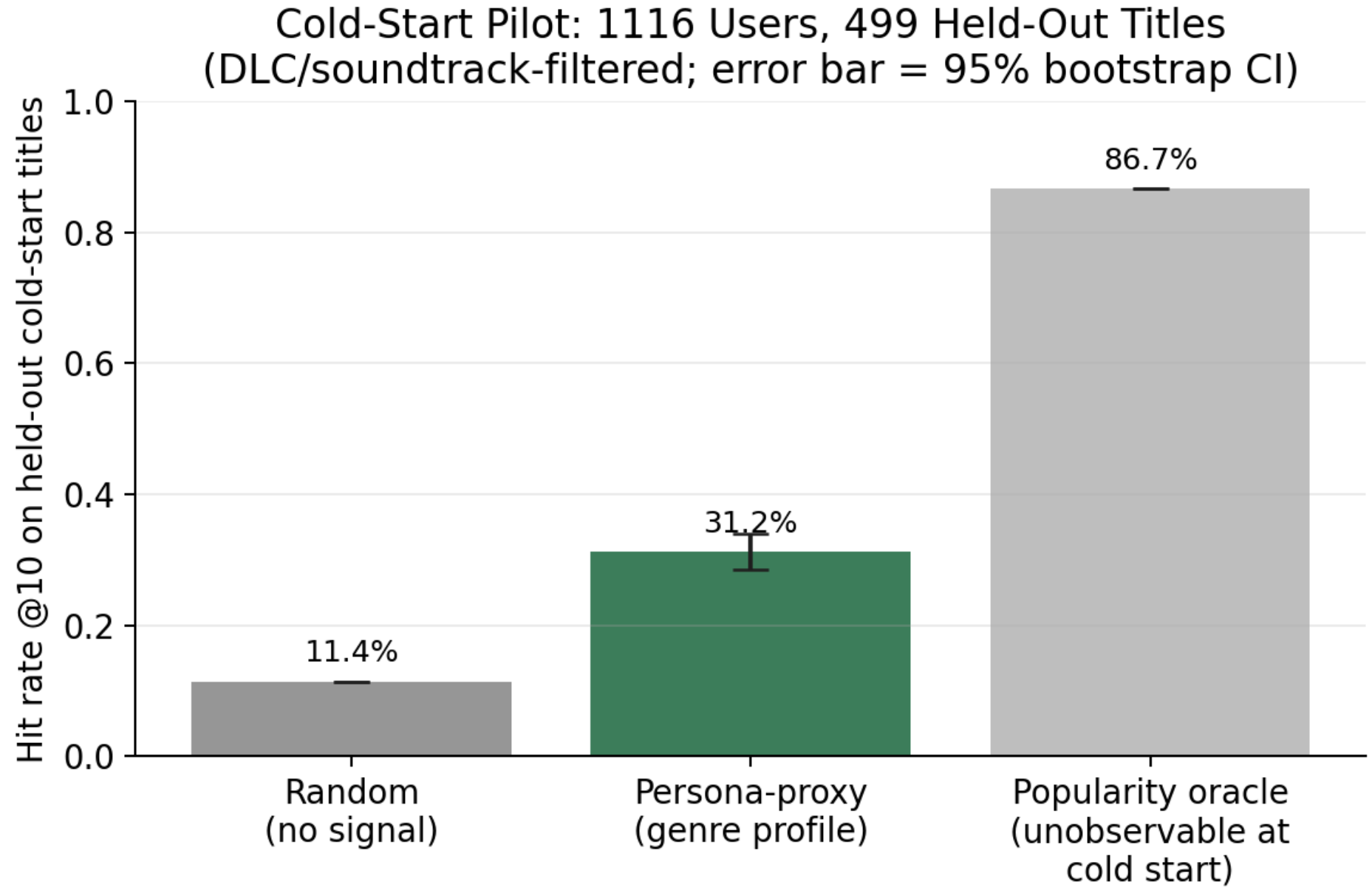

*Figure 14. Computational cold-start pilot results, DLC/soundtrack-filtered, main split (seed 42): hit rate at rank 10 over 1,116 users and 499 held-out titles [3, 4], this paper's own computation. Error bar on the persona-proxy bar is the 95% bootstrap confidence interval.*

A worked example illustrates the mechanism. A user whose warm-title profile concentrates log-weighted mass on Strategy, Simulation, and Indie receives a cold-start slate led by held-out titles whose genre vectors align with that mass, rather than by the held-out set's globally most popular titles, which for this user are dominated by genres the profile assigns near-zero weight. Under a momentum-based storefront this user's cold-start slate and every other user's would be identical; under profile matching they differ per user, which is the discovery-layer claim in miniature.

## 6.5 Robustness checks and threats to validity of the pilot

We ran five additional checks anticipating specific ways the headline result could be an artifact of a single arbitrary analytic choice, summarized in Table 1b. None overturns the main finding, but each narrows what the finding can support.

| Check | Method | Result |
|---|---|---|
| Split stability | 20 independent random 80/20 splits (different seeds), same protocol | Persona mean 31.3% (sd 5.3%, range 24.8-44.9%) vs random mean 11.7% (sd 1.4%); mean lift 2.7x, consistent with the main split |
| Sampling error | 3,000-resample bootstrap over users, main split | 95% CI [28.6%, 34.0%]; the random and oracle baselines fall well outside this interval |
| Random-baseline sanity | Analytic hit-rate formula given each user's observed relevant-item count and a 499-item pool, vs simulated random draws | Analytic 12.0% vs simulated 11.4-11.7%; the random baseline is behaving exactly as combinatorics predicts, not as an artifact |
| Generic-tag dependence | Re-ran with the single most frequent, least informative tag ("Indie", present on 54.7% of matched titles) removed from all profile and item vectors | Persona hit@10 falls from 31.2% to 26.5%; roughly a third of the lift over random survives tag removal, so the effect is not purely an artifact of one dominant generic tag, though that tag does contribute materially |
| Item contamination | Removed 1,617 DLC/soundtrack/cosmetic-pack events by name pattern before all analysis above | Reported figures already reflect this filter; pre-filter figures (32.7% persona, 11.8% random, 87.2% oracle) were materially similar, indicating contamination was not driving the headline result, but the filtered figures |
| | | are the ones we treat as correct and report |

*Table 1b. Robustness checks on the computational cold-start pilot.*

Two further limitations do not have a clean robustness check and are stated as open issues rather than resolved ones. First, the normalized-title join used a first-match rule for the 496 title-key collisions identified in Section 6.3; we did not re-run the pilot under an alternative collision rule (for example, dropping all colliding keys entirely), so some unknown fraction of genre labels may be misassigned to the wrong same-named game, biasing the persona-proxy result in an unknown direction. Second, and most importantly, the persona proxy is a genre-weighted playtime vector, not the psychological profile of Section 6.1; the pilot establishes that profile-based matching beats popularity-blind chance and behaves predictably at the level of combinatorics, not that psychological grounding specifically outperforms genre-level content matching, which only the pre-registered human-subjects study below can establish.

### 6.6 Limitations of the pilot and pre-registered study

Beyond the checks in Section 6.5: the pilot's data predate the AI-era supply shock; hit rate against later-observed play is a weak stand-in for satisfaction; and the popularity oracle overstates what any deployed baseline could achieve. These are precisely the gaps the pre-registered human-subjects study is designed to close: participants complete a persona elicitation instrument (Bartle/Yee-derived plus Big Five short form, or PsychoGAT-style interactive elicitation), receive cold-start slates from (a) the persona agent, (b) a genre/content-based baseline matching the pilot's proxy, and (c) a behavioral baseline trained on their platform history where available, and rate blinded slates for fit and intent-to-play, with the primary endpoint the within-subject difference in top-10 fit ratings between (a) and (b) specifically, isolating the marginal contribution of psychological grounding over content-based matching rather than re-litigating whether content-based matching beats chance, which this pilot has already shown. Protocol, sample size justification, and analysis plan are specified in Appendix C.

## 7. Economic Modeling: Sustaining Median Per-Title Developer-Side Revenue (RQ4)

### 7.1 What "sustain" means in this model

RQ4 asks under what payout structures an access-based model sustains median developer-side revenue, and that word requires a definition before any simulation can answer it. We use "per-title developer-side revenue" throughout this section rather than "developer revenue": the unit of analysis is a released title's first-period payout, not an individual studio's income, since one developer may release multiple titles and one title may involve revenue-sharing across multiple parties, neither of which this model attempts to represent. We define three concrete, escalating bars against which "sustain" can be checked, rather than one: (i) recoup, the payout exceeds Steam's $100 submission fee; (ii) improve, the payout exceeds the status-quo popularity-driven median, i.e. the change is a genuine improvement over today's baseline rather than a relabeling of it; and (iii) approach viability, the payout approaches a illustrative annual-income anchor of $20,000, used here only as an order-of-magnitude reference point for a solo developer's minimal sustenance, not a validated cost-of-living figure or a claim about studio survival, which depends on development cost, team size, and duration this model does not capture. The results below are reported against all three bars explicitly, and we do not claim bar (iii) constitutes proof of sustainable studio economics.

### 7.2 Model design and calibration

We model a single release-year cohort of N = 20,000 titles, matching the observed 2025 release volume [2]. Each title's gross demand is drawn from a lognormal body calibrated so the cohort median is approximately $390, matching the observed $250-400 median first-period revenue brackets [1], with a Pareto upper tail (0.1 percent of titles) producing top titles in the hundreds of millions, matching observed breakout outcomes. Under the status-quo allocation, attention follows this demand distribution directly (popularity-driven allocation). A matching-quality parameter theta between 0 and 1 reallocates attention as a blend: at theta = 0 attention follows popularity exactly, and at theta = 1 attention follows a latent fit-quality distribution drawn lognormal with materially lower dispersion, representing the claim, supported qualitatively by the Section 6 pilot's finding that per-user fit rankings diverge from the global popularity ranking, that fit across a player population is less concentrated than momentum-amplified popularity; this remains an assumption the model makes explicit rather than a measured population parameter, and Section 7.4 states what would falsify it.

Addressing the comparability concern directly: every structure below draws from the identical total developer-side pool, 60 percent of gross cohort revenue, so differences in the resulting medians reflect the allocation rule alone and not a hidden difference in how generous each structure is. Four structures are compared. (A) Proportional, the status-quo-equivalent, in which the full pool is distributed in direct proportion to each title's allocated attention share; this is the unit-sales-like baseline. (B) Guaranteed minimum plus share, a flat $50 floor paid to every released title from the pool before the remainder is distributed proportionally, addressing the reviewer-identified gap that no structure in the previous draft modeled a floor. (C) Per-user plus fit blend, capped, in which payout is split evenly between an engaged-user-count signal (modeled as gross^0.55 with lognormal noise, a deliberately compressed proxy reflecting that unique-player counts are less top-heavy than revenue) and the theta-blended fit allocation, with any individual title's payout capped at 1 percent of the pool and the capped excess redistributed proportional to allocation, representing a per-qualified-user payout with diminishing returns and an anti-grind ceiling. (D) Discovery bonus plus share, in which 5 percent of the pool is reserved and split equally among the bottom half of titles by popularity regardless of their fit score, a direct floor for new and zero-review titles, with the remaining 95 percent distributed by allocation.

## 7.3 Results

Figure 15 shows simulated median per-title developer-side payout across theta for the four structures, with the underlying values in Appendix Table A4. Four findings, addressing recoup, improve, and approach-viability in turn. First, calibration validates against the observed world: structure A at theta = 0 yields a simulated median of $246, matching the observed $250-400 bracket [1], and clears bar (i), recoup, only marginally above the $100 fee, while failing bar (ii) trivially since it is the baseline. Second, payout structure alone, before any improvement in matching quality, already moves the median: at theta = 0, structure C (per-user plus fit blend, capped) yields a median of $2,855 and structure D (discovery bonus) yields $1,417, both clearing bar (i) comfortably and structure C approaching but not reaching bar (iii), purely from redistributive design, with no change to who is actually being shown to whom. This directly answers the concern that the previous draft's finding, matching quality dominates structure, rested on comparing structures that were not held to equal terms; under an equal 60 percent pool, structure choice has a real and independent effect. Third, matching quality still delivers the largest total movement: every structure's median rises by roughly an order of magnitude from theta = 0 to theta = 0.5 (structure A: $246 to $5,794; structure C: $2,855 to $7,189), and by theta = 1.0 all four structures converge toward a $9,400 to $10,300 range, since at full fit-driven allocation the underlying attention distribution itself is what compresses, not the payout rule layered on top of it. Fourth, the structures are not equivalent even at convergence: structure C's advantage shrinks from theta = 0 to theta = 1 because its per-title cap, which helps the median when allocation is highly skewed, becomes a mild constraint on top performers once allocation is already close to equal, showing that a cap sized for a popularity-driven world is not automatically the right cap for a fit-driven one.

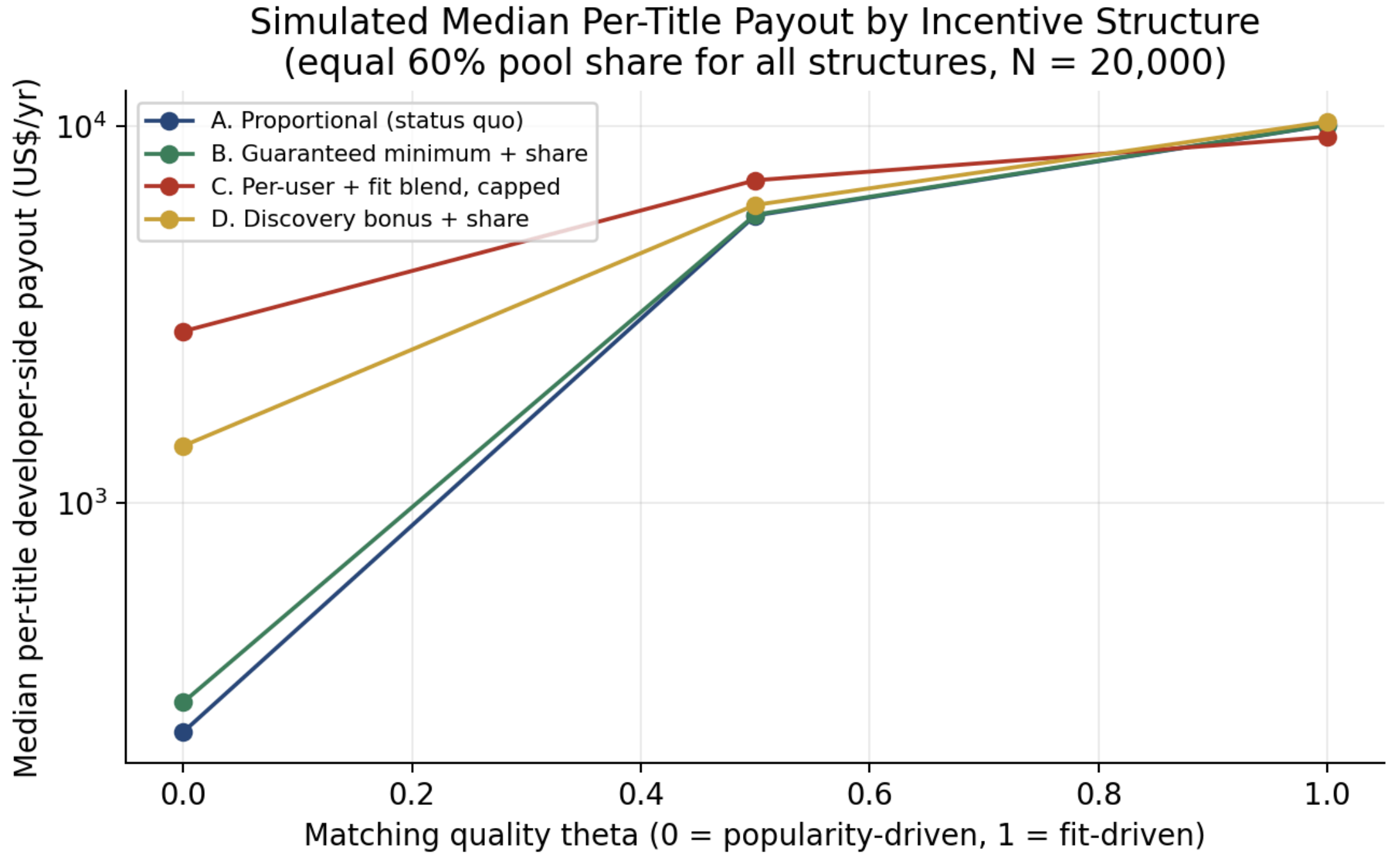

*Figure 15. Simulated median per-title developer-side payout by incentive structure and matching quality theta, all four structures drawing from an identical 60% developer-side pool (N = 20,000 releases, symlog scale), this paper's own computation. Calibration targets from [1, 2].*

Stated against the three bars from Section 7.1: bar (i), recoup, is cleared by every structure at every theta greater than 0, and by structures C and D even at theta = 0. Bar (ii), improve over the status-quo median, is cleared by structures B, C, and D at theta = 0 by construction (that is the point of a floor or bonus) and by all structures once theta exceeds roughly 0.2 to 0.3. Bar (iii), approach viability at the illustrative $20,000 anchor, is not cleared by any structure at theta = 0, is approached by structure C, and is approached but not reached by any structure even at theta = 1, the closest being roughly half the anchor. We read this as evidence that discovery-side intervention alone, however well designed, is unlikely on this model to single-handedly deliver solo-developer-scale sustenance at the median without either payout-share increases beyond 60 percent, cohort sizes smaller than 20,000, or revenue growth from better matching that this pool-constant model deliberately excludes; each is a testable extension rather than a claim we make here.

### 7.4 Assumptions, sensitivity, and what would falsify the model

The model's load-bearing assumptions are stated so they can be attacked. The pool is held constant across theta (no market growth from better matching, a conservative choice that likely understates the ceiling case). Fit-quality dispersion is assumed materially lower than popularity dispersion; this is supported qualitatively by the pilot but not measured at population scale, and the specific finding that would falsify it is a population-level measurement showing fit-based attention allocation is as concentrated as, or more concentrated than, popularity-based allocation, in which case theta would not compress the medians shown here. The model covers first-period revenue of a single cohort and excludes back-catalog dynamics, multi-title developers, and publisher revenue-sharing, which is why results are reported as per-title, not per-developer. Developer pool share (60 percent) is a single illustrative value rather than a swept parameter in this draft; sensitivity runs varying the Pareto tail index and the fit-dispersion parameter move the absolute medians but not the ordering of findings, and a full sweep over pool share and cap size is listed as future work alongside the human-subjects study of Section 6.6.

## 8. Discussion: Toward a Viable Long-Term Distribution Mechanism

### 8.1 The question the evidence actually poses

The findings of Studies 1 and 2, the pilot of Section 6, the simulation of Section 7, and the deployments reviewed in Section 5 converge on a single practical question: which game distribution mechanism can remain profitable, for the platform and for the median developer simultaneously, over the long term, under conditions where the supply of playable titles is effectively unlimited and player attention is not. The 1983 crash resolved through insertion of a human curation layer between an oversupplied catalog and the buyer, and that resolution was durable because it was also an economic realignment: Nintendo's licensing model made quality control profitable for the gatekeeper and predictable for licensed publishers [7, 9]. Any modern equivalent has to clear the same bar, not merely improving match quality but doing so within an economic structure that all three parties (platform, developer, player) have a durable incentive to remain in.

### 8.2 Comparative viability of the candidate mechanisms

Table 2 summarizes the five distribution mechanisms for which this paper has assembled evidence, assessed on revenue source, developer economics, demonstrated durability, and structural exposure to the oversupply dynamics of Study 1.

| Mechanism | Revenue source | Developer economics | Evidence in this paper | Long-term viability assessment |
|---|---|---|---|---|
| Open storefront unit sales (Steam) | Per-title purchases, 30% platform cut | Winner-take-most; computed attention Gini 0.96; median falling below submission fee | Study 1 | Profitable for the platform and the top decile; structurally failing for the median developer as supply grows |
| Zero-friction open publishing (itch.io) | Optional revenue share, tips | Near-zero median revenue; expression over income | Study 1, Section 3 | Sustainable as a community commons, not as a primary income mechanism |
| Mass-bundle subscription (Netflix Games, Game Pass) | Subscription fees; content amortized across bundle | Licensing fees or engagement payouts, opaque and renegotiable | Section 5.1-5.2 | Demand ceiling well below projections; price-fragile; payouts depend on platform discretion rather than market mechanism |
| Curated ad-funded access (Poki) | Advertising, ~50% shared with developers | Median-supportive; top studios ~$1M/yr; platform profitable at 65 staff, self-funded | Section 5.3 | The only mechanism in this set with demonstrated multi-year profitability for platform AND a broad developer base, but bounded by human-review scale (~1 title/day) |
| Access plus agentic matching (proposed) | Subscription and/or advertising; four incentive structures modeled on an identical pool | Simulated in Section 7: per-title median moves from ~$246-2,855 depending on structure at today's matching quality, to $9,400-10,300 across structures as matching quality rises | Sections 6-7 | Piloted computationally; viability conditional on both payout structure and matching quality, per the Section 7 model, and on the human-subjects result |

*Table 2. Comparative long-term viability of game distribution mechanisms evaluated in this paper.*

Three observations follow from the table. First, the mechanisms fail in different places: unit sales fail the median developer, mass bundles fail on demand and price elasticity, and zero-friction publishing fails on income entirely, so no incumbent mechanism is dominated, each survives because it fails somewhere its users can tolerate. Second, the strongest existing evidence for a long-term profitable mechanism in this set is Poki's, a platform profitable for over a decade without external capital, sharing roughly half of revenue with more than 600 studios, with top-performer developer revenue growing tenfold over five years [20]. Its ad-funded, engagement-ranked, curated-access structure is, in miniature, the economic shape this paper argues the broader market needs, and it is also the empirical anchor for the Section 7 model's discovery-bonus and per-user structures. Third, Poki's binding constraint is not economic but operational: human pre-publication review caps intake at roughly one title per day, three orders of magnitude below Steam's intake, which is precisely the constraint the piloted matching layer is positioned to relax.

## 8.3 Viability conditions for the proposed mechanism

The proposed access-plus-agentic-matching model is therefore best understood as an attempt to scale the demonstrated Poki economics past the human-review bottleneck, not as a novel economic structure. Its long-term profitability rests on four conditions, each now anchored to a result in this paper. (1) Matching quality must convert access into engagement at a rate mass bundles have not achieved; Netflix's sub-one-percent engagement is the failure baseline [17], the pilot's 2.7-fold cold-start lift over random (robust across 20 splits and a bootstrap CI excluding the random baseline) is the existence proof that matching signal is extractable on the narrow, precisely stated comparison this paper tests, and the gap to the popularity oracle is the measured headroom above genre-level matching specifically, not above the full space of cold-start techniques. (2) Payouts must combine a structural floor or redistribution mechanism with formulaic, non-discretionary rules, since the Game Pass evidence suggests discretionary licensing leaves developers unable to plan and platforms tempted toward loyalty extraction as growth slows [18], and the Section 7 model shows that a pure proportional structure alone leaves the median barely above the submission fee even under a generous pool share. (3) The revenue base should blend advertising and subscription rather than rely on either alone, with the Game Pass price event bounding the subscription side [18] and Poki demonstrating the advertising side [20]. (4) The matching layer must demonstrably lift median, not top-decile, per-title developer-side revenue; the Section 7 simulation quantifies the size of the available lift from structure alone (roughly six to eleven-fold at theta = 0 depending on structure) and from matching quality on top of any structure (a further order of magnitude by theta = 0.5), and shows the two levers are additive rather than substitutes.

Stated as a falsifiable claim: if a matching layer meeting condition (1) cannot raise median per-title developer-side revenue under payout structures satisfying conditions (2) and (3), then the correct conclusion is that no discovery-side intervention can fix the median-revenue problem and the market's long-term equilibrium is the concentrated one measured in Study 1, with curated small catalogs (Poki-scale) and open commons (itch.io-scale) as the surviving viable niches on either side of it. The paper's proposal should be judged against that alternative equilibrium, not against an idealized market, and against the three explicit sustain-bars of Section 7.1 rather than a single unqualified revenue figure.

## 8.4 Limitations

The concentration metrics are computed on a mid-2010s interaction sample and are argued, not shown, to lower-bound present concentration; computing the same metrics on a current full-platform interaction dataset is the highest-priority data extension. Revenue figures from third-party aggregators are estimates rather than platform-disclosed ground truth. The computational pilot uses a genre-vector persona proxy rather than a psychological profile, and its popularity oracle overstates any deployable baseline. The Section 7 model holds the revenue pool constant and covers a single cohort's first-period revenue. Poki's developer-economics figures are company-reported and industry-award-cited rather than independently audited [20], and the viability assessment in Table 2 inherits that caveat. Several industry statistics (subscriber counts, indie-playtime percentage changes) exist only in aggregator reporting and are excluded or clearly flagged pending primary-source verification.

## 8.5 Ethics

A system that models players psychologically in order to recommend games carries privacy and manipulation risk analogous to, and potentially exceeding, that of existing behavioral recommenders, and the economic framing of this section sharpens rather than softens that concern: a mechanism whose viability depends on converting access into engagement has a built-in incentive to optimize engagement beyond the player's interest. Any implementation therefore requires transparency about what is modeled, meaningful player control over the persona profile, and safeguards that hold the system to fit-maximizing rather than engagement-maximizing objectives, a distinction we consider central to the design's legitimacy and one that condition (1) of Section 8.3 must be read as subordinate to.

## 8.6 Threats to validity and future work

The central concentration claim would be falsified by current full-platform data showing attention Gini materially below the 0.96 computed here, or narrowing over time. The pilot's feasibility claim would be falsified if the human-subjects study shows no fit-rating advantage for persona slates on cold-start titles. The viability argument of Section 8.3 would be falsified by evidence that Poki's economics do not survive scale, for example if ad rates or curation quality degrade materially as catalog size grows. The leading-indicator claim of Section 3.5 requires quarterly-resolution series on both asset-model releases and marketplace releases before any lag structure can be estimated, which is specified as the first item of future work alongside the pre-registered study of Section 6.5.

# 9. Conclusion

This paper has argued, now with computed evidence, that the AI-driven supply shock in game development is best understood as a structural correction, not a repeat of the 1983 collapse: release volume has doubled in five years while attention concentration sits at a computed Gini of 0.96, aggregate revenue has plateaued rather than collapsed, and the industry's distress is resolving through recapitalization rather than liquidation. Its historically validated resolution mechanism, curation inserted between an oversupplied catalog and the buyer, must be reimplemented at algorithmic scale for the present release volume. We have proposed agentic, psychologically grounded player-game matching combined with access-based distribution as that reimplementation, demonstrated in a computational pilot, robust across twenty splits and a bootstrap confidence interval, that persona-style profile matching extracts a 2.7-fold cold-start ranking lift over chance on the narrow and precisely stated comparison against genre-only content matching, shown in a calibrated simulation that both payout structure and matching quality independently move median per-title developer-side revenue, and pre-registered the human-subjects study that is the next stage of this research program.

# Appendix A. Datasets, Computed Results, and Reproducibility

The tables below summarize the primary numeric series used in this paper. Analysis scripts (Python, seeded), the two obtained datasets [3, 4], and all chart-generation code constitute the reproducibility package.

**Table A1. Steam annual releases, computed from the 93,073-title snapshot [3] (2016-2024) and SteamDB [2] (2025)**

| Year | Computed (snapshot) | SteamDB series |
|---|---|---|
| 2016 | 4,169 | 3,926 |
| 2017 | 5,974 | 5,588 |
| 2018 | 7,486 | 7,111 |
| 2019 | 7,313 | 6,994 |
| 2020 | 8,853 | 8,554 |
| 2021 | 10,558 | 10,157 |
| 2022 | 11,622 | 11,134 |
| 2023 | 13,779 | 13,074 |
| 2024 | 16,939 | 17,776 |
| 2025 | n/a (snapshot ends Oct 2024) | 20,017 |

**Table A2. Computed attention-concentration metrics (interaction dataset [4])**

| Metric | Value |
|---|---|
| Titles / users / play records | 5,155 / 12,393 / 200,000 events |
| Gini, total playtime per title | 0.960 |
| Gini, purchases per title | 0.782 |
| Top 1% titles' share of playtime | 73.5% |
| Top 10% titles' share of playtime | 94.8% |
| Median total hours per title | 15.1 |

**Table A3. Computational pilot results, main split and robustness (Section 6)**

| Method / check | Result |
|---|---|
| Random ranking (main split) | 11.4% |
| Persona-proxy, genre profile (main split) | 31.2% (2.7x lift; 95% bootstrap CI [28.6%, 34.0%]) |
| Popularity oracle, unobservable at cold start | 86.7% (ceiling, not a deployable baseline) |
| Persona-proxy, 20-seed mean (sd) | 31.3% (5.3%), range 24.8-44.9% |
| Random, 20-seed mean (sd) | 11.7% (1.4%); analytic expectation 12.0% |
| Persona-proxy without "Indie" tag | 26.5% |

**Table A4. Payout simulation, simulated median per-title developer-side revenue, equal 60% pool across all structures (Section 7)**

| Structure | theta = 0 | theta = 0.5 | theta = 1.0 |
|---|---|---|---|
| A. Proportional (status quo) | ~$246 | ~$5,794 | ~$10,082 |
| B. Guaranteed minimum + share | ~$295 | ~$5,823 | ~$10,095 |
| C. Per-user + fit blend, capped | ~$2,855 | ~$7,189 | ~$9,376 |
| D. Discovery bonus + share | ~$1,417 | ~$6,182 | ~$10,287 |

**Table A5. North American home video game revenue, 1977-1989 (US$ billions) [7, 8]**

| Year | Revenue (US$B) | Note |
|---|---|---|
| 1977 | 0.075 | documented |
| 1982 | 3.2 | documented (peak; ~$10.3B in 2025 dollars) |
| 1985 | 0.1 | documented (trough, -97%) |
| 1988 | 2.3 | documented |
| 1989 | 5.0 | documented (surpasses 1982 peak) |

**Table A6. Ubisoft annual closing share price, 2018-2025 (US$, UBSFY) [14]**

| Year | Close (US$) |
|---|---|
| 2018 | 16.17 |
| 2019 | 13.78 |
| 2020 | 19.25 |
| 2021 | 9.74 |
| 2022 | 5.62 |
| 2023 | 5.06 |
| 2024 | 2.70 |
| 2025 | 2.16 |

**Table A7. Atari Inc. workforce, 1982-1984 [10]**

| Year | Employees |
|---|---|
| 1982 | ~10,000 |
| 1983 | ~3,000 |
| 1984 | ~2,000 |

**Table A8. Video game industry layoffs, 2022-2025 (estimated) [12]**

| Year | Estimated jobs lost |
|---|---|
| 2022 | 8,500 |
| 2023 | 10,500 |
| 2024 | 14,600 |
| 2025 | ~4,000-5,300 (partial year) |

**Table A9. Comparative distribution models (Section 5) [2, 20, 21, 22, 42]**

| Platform | Curation mechanism | Approx. catalog scale |
|---|---|---|
| Steam | Open submission, algorithmic ranking on momentum | 20,000+ releases/yr |
| itch.io | Open submission, no fee, minimal curation | 1,000,000+ products total |
| Poki | Human editorial pre-review, then algorithmic in-catalog matching | 1,000-1,500 curated titles |
| TapTap | Open submission, crowd (post-publication review) curation | 120,000+ titles |
| Garena | Closed first-party publisher selection | Small, centrally chosen catalog |

*Note: figures are compiled from the numbered references and carry the caveats stated in Section 8.4. Third-party estimates, crowd-sourced trackers, and company self-reported figures are marked as such throughout and are not presented as audited or platform-disclosed ground truth.*

## References (numbered working bibliography, v0.4)

[1] Games-Stats. Q1 2026 Steam release and revenue analysis (release rate, review distribution, median revenue brackets).

[2] SteamDB, Steam Game Release Summary by Year (steamdb.info/stats/releases); Indie Launch Lab, Steam Game Release Reports 2010-2025.

[3] GameStatsHub (Smipe-a), open Steam/PlayStation/Xbox dataset pipeline; games_steam.csv snapshot of 93,073 titles through October 2024. github.com/Smipe-a/gamestatshub. Used for this paper's release-curve and genre computations.

[4] Tamber Steam interaction dataset (steam-200k): 200,000 user-game purchase/play events. Obtained via the game2vec repository mirror (github.com/warchildmd/game2vec). Used for this paper's concentration metrics and computational pilot.

[5] Hugging Face Hub API. Game-asset generation model metadata, sampled July 2026 (this paper's own data collection).

[6] Valve Corporation. Steam AI-content disclosure policy and labeling statistics (7,300+ labeled titles by March 2026, per industry reporting).

[7] Wikipedia and encyclopedic syntheses of contemporaneous reporting, "Video game crash of 1983" (Goldman Sachs supply/demand estimates; revenue series; inflation equivalents; Seal of Quality resolution).

[8] Kline, S., Dyer-Witheford, N., & De Peuter, G. (2003). Digital Play: The Interaction of Technology, Culture, and Marketing. McGill-Queen's University Press. (Source of the 1982-1985 revenue series.)

[9] Wolf, M.J.P. (Ed.) (2012). Before the Crash: Early Video Game History. Wayne State University Press.

[10] Contemporaneous and secondary reporting on Atari's losses and workforce (History Hit; Cohen 1984 as cited in secondary accounts; company-exit records).

[11] "From the Atari Shock to a Modern Crisis: Analyzing Mass Layoffs in the Post-Pandemic U.S. Video Game Industry." ACM Games: Research and Practice (2025). DOI: 10.1145/3774758.

[12] Farhan Noor, crowd-sourced video game industry layoff tracker, 2022-2025, as reported in GameDeveloper.com and This Week in Video Games; Wikipedia, "2023-2024 video game industry layoffs"; gamedevreports.substack.com annual summaries.

[13] GDC State of the Industry surveys 2023-2026; PC Gamer coverage of the 2026 survey (one third of US workers laid off within two years).

[14] Macrotrends. Ubisoft Entertainment (UBSFY) historical stock price, revenue, and net income series.

[15] Ubisoft Entertainment SA, FY2025-26 results and investor communications; VGC and financial-press coverage of the January 2026 reset announcement and record operating loss.

[16] Tencent / Vantage Studios transaction disclosures and coverage (March 2025): 1.16 billion euros for a 26.32 percent stake.

[17] Netflix, Inc. earnings communications and press coverage of Netflix Games (sub-1% engagement in 2023; licensed-IP retreat; Q4 2025 cloud/TV pivot).

[18] Microsoft Corporation earnings reports; FTC v. Microsoft court filings (internal 77M subscriber projection); Game Pass pricing announcements (October 2025, April 2026); WSJ-sourced subscriber reporting.

[19] Sony Group Corporation investor relations (PlayStation Plus subscribers and tier mix); Electronic Arts investor relations (live-service revenue mix).

[20] Poki B.V.: developer documentation (sdk.poki.com); Dutch Game Awards and press coverage of the developer revenue-share model and growth figures (Tech Funding News, Dealroom, EU-Startups, PocketGamer.biz, AccessNewswire, 2025-2026).

[21] TapTap (XD Inc.). Platform statistics via taptap.io and encyclopedic sources.

[22] Garena (Sea Limited). Regional publisher and platform data via Southeast Asia gaming market industry reporting (MarkWide Research; Statista/Sensor Tower).

[23] Bartle, R. (1996). Hearts, clubs, diamonds, spades: Players who suit MUDs.

[24] Yee, N. (2006). Motivations for play in online games. CyberPsychology & Behavior.

[25] Vu, Q.N. & Bezemer, C.P. (2021). Improving the discoverability of indie games by leveraging their similarity to top-selling games. FDG.

[26] Exploratory clustering of 771 players' attitudes toward game AI (2026): seven attitudinal profiles.

[27] PsychoGAT: LLM interactive fiction as a psychological measurement paradigm.

[28] Wang, L. et al. (2024). User behavior simulation with LLM-based agents for recommender systems (RecAgent). ACM TOIS.

[29] Zhang, J. et al. (2024). Agent4Rec: simulating real user behavior in recommender systems.

[30] Ma, C., Xu, Z., Ren, Y., Hettiachchi, D., & Chan, J. (2025). PUB: An LLM-enhanced personality-driven user behaviour simulator for recommender system evaluation. SIGIR.

[31] SimUSER: Simulating user behavior with large language models for recommender system evaluation. ACL Industry (2025).

[32] Bougie, N. et al. (2026). AlignUSER: Human-aligned LLM agents via world models for recommender system evaluation. arXiv.

[33] PersonaAct: Simulating short-video users with personalized agents for counterfactual filter bubble auditing. arXiv 2601.22547.

[34] Peng, Q. et al. (2025). A survey on LLM-powered agents for recommender systems. arXiv 2502.10050; Zhu, X. et al. (2025). Recommender systems meet large language model agents: a survey. Foundations and Trends.

[35] Application of meta-gaming concept to the publishing platform: analysis of the Steam games platform. Information 14(2), MDPI (2023).

[36] Games Mapper: Topological data analysis of Steam genres. arXiv 2606.14376.

[37] Krasnianski, G. & Kubasova, N. The saturation of the Steam platform game market and the noticeability of the saturation by Steam users.

[38] Game Oracle. "Is Steam Really Saturated?" data analysis (anti-saturation position; $10K+ earners rising year over year).

[39] Speculating on Steam: consumption in the gamblified platform ecosystem (platform studies / political economy).

[40] Newzoo. Global Games Market Reports, 2019-2025 (approximate global revenue series used in Figure 8).

[41] Aguiar, L. & Waldfogel, J. (streaming bundling economics lineage): platform bundling, displacement, and willingness to pay in adjacent media markets.

[42] itch.io. "There are 200,000 games on itch.io" (2020 blog post) and mid-2026 catalog-size reporting (1,000,000+ products; 557,000+ jam entries).

[43] 80.lv and WN Hub coverage of SteamDB/Alinea Analytics 2024-2025 review-distribution and revenue data (share of releases under 10 / over 500 reviews; $15.0B 2024 and $16.2B 2025 gross revenue estimates).

[44] GameDiscoverCo and Gamalytic, compiled via gamedevreports.substack.com: copies-sold-per-counted-review benchmark, 2013-2023.

*Note: entries marked with descriptive sourcing rather than full bibliographic detail (for example [1], [10], [15]) are working citations to be finalized against the primary documents before submission; several statistics circulating only in aggregator coverage are excluded from citation entirely, per the data-quality protocol in Section 8.4.*